\documentclass[%
 aip,
 reprint,
 amsmath,amssymb,
 floatfix,
]{revtex4-2}
\usepackage{graphicx}
\usepackage{xcolor}
\usepackage{bm}
\usepackage{siunitx}
\usepackage{array}
\usepackage{booktabs}
\usepackage{tabularray}
\usepackage{textcomp}
\newcommand{\blue}{\textcolor{black}}

\begin{document}

\title{Writing with Near-Field: towards Atomic Resolution in Material Processing}

\author{
Haoran Mu$^1$, Hsin-Hui Huang$^1$, Tomas Katkus$^1$, Darius Gailevi\v{c}ius$^2$, Saulius Juodkazis$^{1,2}$}
\email{sjuodkazis@swin.edu.au}
\affiliation{Optical Sciences Centre, 
Swinburne University of Technology, Hawthorn, VIC 3122, Australia}
\affiliation{Laser Research Center, Physics Faculty, Vilnius University, Saul\.{e}tekio Ave.\ 10, 10223 Vilnius, Lithuania}

\date{\today}

\begin{abstract}
Direct laser writing with ultra-short pulses is the most flexible route to two- and three-dimensional (2D/3D) material modification, yet its resolution is habitually judged against the optical diffraction limit. This review argues the opposite: the finest and most reproducible modifications are obtained not by shortening the wavelength but by localising the \emph{absorber}. Through the threshold effect and the formation of a transient epsilon-near-zero \blue{(ENZ, $\varepsilon'\!\to\!0$) state - the permittivity of the transient dielectric-to-metal (``Die-Met'') 
state of the excited solid - } the deposited energy is confined to a dipole-like nano-seed whose near-field intensity falls steeply with distance $r$ (as $r^{-6}$ for an ideal point dipole). Hence, the localisation is set by the seed size rather than by the focal wavelength. \blue{The final feature is additionally shaped by the material response: post-pulse heat flow, cavitation and quenching set how faithfully the modification follows the deposited-energy profile.} We further show that \emph{polarisation} controls the directionality and anisotropy of the energy deposition: the quasi-static near-field and the interface boundary conditions select the writing direction and the orientation of the inscribed modification, down to the scale of a single inscribed defect. We critically survey energy-deposition mechanisms from \SI{100}{nm} to the single-nanometer scale and connect them to engineering that already works: stitch-free large-area patterning and record aspect-ratio ($\sim\!10^3$--$10^4$) dicing and wafering of transparent crystals. Finally, we advance a conjecture of guided volumetric energy deposition, in which the laser-defined/inscribed lossy nano-seeds act as antennas and heaters that let cavity microwaves anneal the laser-modified nano-volumes from within, much as a \SI{2.45}{GHz} oven heats food volumetrically through water-dipole absorption. \blue{Throughout, ``atomic resolution'' is used in the quantum-confined sense defined in the Introduction and denotes a direction of travel: 10--20\,nm features are demonstrated experimentally, whereas deterministic single-defect and single-bond modification remain research goals.} The defensible thesis throughout is that ultrafast direct write beats diffraction-limited intuition because the material itself forms a transient nanoscale absorber whose near-field and boundary conditions govern the subsequent energy deposition.
\end{abstract}

\maketitle
\tableofcontents

\section{Introduction: the two Moore's laws}\label{sec:intro}

The electronics (e-)Moore's law of doubling the transistor count roughly every two years~\cite{Moore1965} has driven computing, telecommunications and, most recently, artificial intelligence, and is built on \emph{parallel}, planar 2D nanolithography. Its optical (o-)Moore's counterpart describes the doubling of the average power (photons per unit time) of ultra-short pulsed lasers \blue{(USPL)}, a trend that has held for the past $\sim\!25$~years~\cite{Han2021,Dubietis2023}. The fundamental distinction between the two laws is that o-Moore's scaling powers \emph{sequential} direct laser writing (DLW) and material processing, whereas e-Moore's success rests on parallelism~\cite{Juodkazis2026stitch}. The question this review addresses is when, and by what physics, sequential direct write can challenge parallel-mode lithography and how the energy of an ultra-short pulse
can be localised to a nanoscale volume that is essentially independent of the illuminating wavelength.

The state of the art in commercial lithography is achieved with \SI{193}{nm} (ArF) water-immersion and \SI{13.5}{nm} extreme-ultraviolet (EUV) sources. EUV
requires vacuum enclosures, reflective masks, and chemically amplified resists whose stochastic line-edge roughness limits resolution~\cite{Itani2013}, and
entails a cost of order \$700~million per node change. Crucially, the photomasks themselves are written by electron-beam lithography (EBL), that is, in a \emph{direct-write}, sequential mode. The energy efficiency of projection tools is poor: $\sim\!\SI{1}{MW}$ of wall-plug power, used to ablate Sn micro-droplets with a CO$_2$ laser, yields only $\sim\!\SI{0.5}{W}$ of useful exposure on the
wafer an end-to-end efficiency of $\sim\!10^{-6}$, while a planar CMOS flow requires $\sim\!10^{3}$ processing steps per wafer. These realities motivate a
paradigm shift toward maskless, energy-efficient, 3D-capable fabrication a ``3D-printing-like'' democratisation of high-end micro- and nano-electronics, photonics and large-area patterning for photovoltaics~\cite{Juodkazis2026stitch,Maksimovic2022}.

At the leading edge of semiconductor manufacturing, lithographic resolution is no longer the only limiting variable; the rate at which that resolution can be deployed has become equally consequential. EUV is required for selected critical layers in approximately 5-nm-class and smaller technologies, yet its availability is concentrated in a single commercial scanner supplier and a highly specialised subsystem network, including sole-source dependencies in critical projection optics.\cite{Kleinhans2025SemiconductorValueChain,Cerutti2023SemiconductorsEU,ASML2025AnnualReport} The transition from 0.33-NA EUV to 0.55-NA High-NA EUV is intended to extend single-exposure scaling and reduce reliance on increasingly complex multipatterning, but industrial insertion requires the scanner, source, mask, resist, metrology, overlay and process-integration ecosystems to mature together.\cite{Rasmussen2023EUVWorkingGroup,Levinson2022HighNA} This qualification burden creates a substantial interval between demonstrated optical resolution and broadly replicated manufacturing capability. The resulting bottleneck is therefore not simply the minimum printable pitch, but the limited rate at which a complete, capital-intensive lithographic ecosystem can be qualified and distributed across the manufacturing base.

Two requirements define the frontier. First, \emph{feature size}: a single modification as small as $\sim\!\SI{10}{nm}$ approaches the ``atomic'' or quantum scale, since a single dopant atom in a crystal, e.g.\ a P atom in Si, behaves as a quantum dot of $\sim\!\SI{5}{nm}$ diameter~\cite{Tabe2010,Wang2024}. We use ``atomic/quantum scale'' in this quantum-confined sense, distinct from atom-by-atom placement, and throughout separate feature size, placement/pitch, \emph{defect determinism} and atomic selectivity, reserving ``atomic'' for cases with single-defect or single-bond evidence. \blue{By this usage, true atomic resolution in the crystallographic sense ($\sim\!\SI{0.1}{nm}$) remains beyond optical direct write; what is within reach, and demonstrated, is the quantum-confinement scale of $\sim\!5$--$\SI{10}{nm}$ together with single-defect photophysics~\cite{Wang2024}.} Second, \emph{resolution} between two adjacent features, which sets the packing density. Both are required for basic quantum devices (single-photon emitters, single-electron transistors) and both, this review argues, are governed by the spatial localisation of energy deposition rather than by the wavelength alone.

The remainder of the paper is organised as a sequence of self-contained topics: far-field focusing (Sec.~\ref{sec:farfield}), the threshold effect (Sec.~\ref{sec:threshold}), a critical survey of deposition mechanisms (Sec.~\ref{sec:survey}), the dipole/ENZ near-field that unifies 2D and 3D writing (Sec.~\ref{sec:dipole}), the role of polarisation relative to the scan direction (Sec.~\ref{sec:polar}), engineering that already works (Sec.~\ref{sec:engineering}), and a new laser-seeded microwave paradigm (Sec.~\ref{sec:paradigm})\blue{, the latter explicitly flagged as a forward-looking, hypothesis-level element of this 
review-perspective}.

\section{Far-field focusing and the diffraction barrier}\label{sec:farfield}

For propagating (far-field) light the Heisenberg uncertainty relation $\Delta x\,\Delta p_x\gtrsim\hbar$ (with $\hbar$ the reduced Planck constant and $\Delta x,\Delta p_x$ the position and transverse-momentum spreads) translates into
the waist-divergence relation $\Delta\theta\sim\lambda/\Delta x$ ($\lambda$ the
wavelength), fixing the trade-off between the lateral
waist $w_0$ and the depth of focus (DoF)~\cite{Li2024} [Fig.~\ref{f-Heis}(a)]. A diffraction-limited spot
is $w_0\simeq 0.5\,\lambda/\mathrm{NA}\sim\SI{1}{\micro\meter}$ for near-infrared pulses at moderate numerical aperture (NA), and the aspect ratio
$\mathrm{DoF}/w_0\sim 2\lambda/(\pi\,\mathrm{NA}^2)$ is therefore bounded: lateral resolution and aspect ratio cannot be improved independently. A Bessel beam,
synthesised with an axicon, replaces the short DoF with a long, non-diffracting core, raising the achievable aspect ratio while leaving the same uncertainty trade-off in force in the transverse plane~\cite{Ganguly2024}.

The essential point for this review is that the focal spot is only the envelope within which energy is delivered. The size of the modified region is set not by $w_0$ but by where the deposited energy density crosses a material threshold [Fig.~\ref{f-Heis}(b)], a distinction that the diffraction limit, framed for linear imaging, does not capture. Two further features of a tight focus matter below. First, at high-NA the focus is non-paraxial and acquires a longitudinal field
component $E_z$ along the propagation axis, opening a resonant-absorption channel that is absent in the paraxial picture. Second, the same nonlinearity that
sharpens the deposited dose also makes the process deterministic: only the peak of the Gaussian, where the intensity $I$ exceeds the modification-threshold intensity $I_{\rm th}$, is written.

\section{The threshold effect: a tens-of-nm seed from a micron focus} \label{sec:threshold}
The USPL modification is governed by a strongly nonlinear energy-deposition rate $W(I)\propto I^{k}$, with effective order $k\sim6$--$10$\blue{. The microscopic channels behind this phenomenological exponent were established in the classical analyses of photoionisation by Keldysh~\cite{Keldysh1965} and of laser-induced breakdown by Bloembergen~\cite{Bloembergen1974}: the relative weight of multiphoton and tunnelling ionisation is set by the Keldysh parameter $\gamma=\omega\sqrt{m^\ast E_g}/(eE)$ (bandgap $E_g$, field amplitude $E$), which is of order unity at the $10$--$100\,\mathrm{TW/cm^2}$ intensities relevant here, so the two channels coexist rather than compete cleanly. The carriers they seed are then multiplied by avalanche (impact) ionisation on the trailing edge of the pulse. The exponent $k$ is therefore an effective, intensity-range-dependent slope of the deposition rate, not a photon count.} For a Gaussian profile $I(r)=I_0\exp(-2r^2/w_0^2)$ (peak intensity $I_0$, $1/e^2$ radius $w_0$, radial coordinate $r$) and on-axis modification threshold $I_{\rm th}$,
\blue{a threshold on the deposited dose $D\propto I^{k}(r)$ is equivalent to a threshold on the local intensity, because $I^{k}$ is a monotonic function of $I$. The modified radius is therefore}
\begin{equation}
a \;=\; \frac{w_0}{\sqrt{2}}\,\sqrt{\ln\!\big(I_0/I_{\rm th}\big)} .
\label{eq:threshold}
\end{equation}

\noindent \blue{Equivalently, $a=(w_0/\sqrt{2k})\sqrt{\ln(D_0/D_{\rm th})}$, where $D_0/D_{\rm th}=(I_0/I_{\rm th})^{k}$ is the peak-to-threshold dose ratio. Expressed through the experimentally controlled ratio $I_0/I_{\rm th}$ (equivalently $E/E_{\rm th}$), the radius therefore does not depend on the nonlinear order $k$. The nonlinearity acts elsewhere: it narrows the dose profile $\exp(-2kr^2/w_0^2)$ itself, so the dose gradient at the modification contour is $k$ times steeper (a sharper, more deterministic edge) and no dose is deposited outside the focal volume. This is consistent with the single-shot ablation study of Garcia-Lechuga \emph{et al.}~\cite{GarciaLechuga2020}, who showed experimentally, across bandgaps of 3.8 to 8.8\,eV and wavelengths of 515 and 1550\,nm, that ablation craters map the beam contour at a strict threshold intensity irrespective of the nonlinear order. For $w_0=\SI{0.5}{\micro\meter}$, Eq.~(\ref{eq:threshold}) gives $a\approx\SI{225}{nm}$ at $I_0=1.5\,I_{\rm th}$ and $a\approx\SI{78}{nm}$ at $I_0=1.05\,I_{\rm th}$, for any $k$. A radius of \SI{30}{nm} requires $I_0$ to be within $0.7\%$ of $I_{\rm th}$, and \SI{10}{nm} within $0.08\%$.}
The same logarithmic dependence simultaneously reduces the depth of the modified volume [Fig.~\ref{f-Heis}(b)]. Deterministic ablation can therefore reach the tens-of-nanometre regime in both lateral and axial directions~\cite{Liu1997,Joglekar2004,Kumar2014}\blue{, provided that the pulse energy and focus are stable at the $\sim\!1\%$ level or better}. The energy required is modest: any phase transition: melting, evaporation, ionisation is induced with
\SIrange{0.01}{10}{eV} per atom~\cite{Gamaly2018}. This is the first key idea of the review: the relevant length scale is set by where the deposited energy exceeds a material threshold, not by $w_0$.

\blue{
The threshold and near-field analyses of this review localise the deposited energy; the final modification is additionally broadened by post-pulse heat flow during electron-lattice relaxation and resolidification. A schematic scaling is
\begin{equation}
w_{\rm final}\;\approx\;2\sqrt{a^2+4D\tau_{\rm q}},
\label{eq:wfinal}
\end{equation}
with $D$ the thermal diffusivity and $\tau_{\rm q}$ the time over which the seed stays above the modification temperature (bounded from below by the electron-phonon relaxation time of $\sim\!1$--$10$\,ps). In fused silica ($D\approx\SI{8.3e-7}{m^2/s}$) heat spreads only $\sqrt{4D\tau}\approx\SI{6}{nm}$ in 10\,ps and $\sim\!\SI{60}{nm}$ in 1\,ns: a 10-nm deposited profile survives as a 10-20\,nm feature provided quenching is sub-nanosecond, which confined microexplosion provides ($\sim\!10^{14}$\,K/s~\cite{Juodkazis2006,Bhuyan2017}). In crystalline silicon ($D\approx\SI{8.8e-5}{m^2/s}$) the same 10\,ps already gives $\sim\!\SI{60}{nm}$ of spreading, which is why the finest demonstrated features in Si rely on thin films, interfaces or stress/cavitation-mediated (rather than purely thermal) transformation~\cite{Lin2020NL}. Localising the deposition is therefore necessary but not sufficient: the quench rate decides how much of the localisation survives, and Eq.~(\ref{eq:wfinal}) gives the practical guidance: low-$D$ hosts (glasses) hold nanoscale features robustly, high-$D$ hosts (Si, metals) require confinement or ultrafast mechanical transformation.}

The threshold effect alone, however, only carries the modification to $\sim\!\SI{30}{nm}$ before the logarithm in Eq.~(\ref{eq:threshold}) saturates and
shot-to-shot fluctuations dominate the edge definition. Reaching the single-nm scale requires a second mechanism that concentrates the field after the seed is formed, the dipole/ENZ near-field of Sec.~\ref{sec:dipole}.

\section{Energy deposition from \SI{100}{nm} to the single-nanometer scale: a critical survey}\label{sec:survey}

We now scout the principal mechanisms by which energy is deposited into matter on scales from \SI{100}{nm} to \blue{the single nanometer}. Table~\ref{tab:survey} summarises the approaches, their demonstrated localisation, the underlying mechanism, and, most importantly, their dependence on wavelength. The trend we emphasise is that the finest, most wavelength-insensitive localisation is obtained when energy is deposited through a \emph{near-field} or absorber-defined mechanism rather than by linear far-field absorption. This is not a claim of universal superiority. Competing high-resolution routes include electron- and ion-beam lithography and milling (EBL, FIB, and He/Ne-ion methods), electron-beam-induced deposition, plasmonic-antenna and near-field scanning optical lithography, microsphere-assisted structuring, AFM thermochemical or local-oxidation writing, scanning-tunnelling atom manipulation, and directed block-copolymer self-assembly. Each route wins on particular metrics. Recent surveys map this competitive landscape~\cite{Lin2025}\blue{, as does the dedicated perspective on ultrafast-laser volume nanostructuring by Stoian~\cite{Stoian2023}}, while microsphere-assisted schemes have reached $\lambda/27$ feature sizes in ambient air~\cite{Yin2025}. \blue{Several far-field demonstrations now sit firmly in the 10-nm class and must be part of any balanced survey: dual-beam overlapping with orthogonal polarisations has produced $\sim\!10$--$12$\,nm ($\lambda/66$) features on Si surfaces in ambient air~\cite{Lin2020NL}; cavity-seeded evanescent-field structuring with Bessel beams has reached 7\,nm ($<\lambda/100$) trenches of micrometre depth in glass~\cite{Zhang2025UFS}; $\lambda/73$ features were written in van der Waals NbOI$_2$ films~\cite{Guan2025}; and nanohole-clad waveguides with depth-to-diameter ratios up to $5\times10^4$ have been written in single crystals~\cite{Fan2026LAM}. Notably, the mechanisms invoked in these works - self-generated near-fields seeded by voids or cavities, thermal-stress localisation, and interference of incident with back-scattered light - overlap with, but are not identical to, the ENZ/dipole picture developed below. Sec.~\ref{sec:alt} weighs these alternatives explicitly.} Our argument concerns the regime in which a transient, self-formed absorber governs the resolution.

\begin{table*}[t]
\caption{Critical survey of energy-deposition methods from \SI{100}{nm} to the single-nanometer/atomic scale. ``$\lambda$-dependence'' indicates how strongly the achievable localisation is tied to the illuminating wavelength.}
\label{tab:survey}
\footnotesize
\begin{tblr}{
  width=\linewidth,
  colspec={X[2.9,l]X[2.3,l]X[4.8,l]X[2.2,l]X[1.4,l]},
  colsep=4pt, rowsep=1.2pt,
  hline{1,Z}={1pt,solid}, hline{2}={0.4pt,solid},
}
Approach & Localisation & Deposition mechanism & $\lambda$-dependence & Refs. \\
193\,nm immersion photolithography & $\sim$38\,nm half-pitch & linear absorption in resist; phase + immersion & strong ($\lambda/\mathrm{NA}$) & \cite{Itani2013}\\
EUV (13.5\,nm) projection & $\sim$13\,nm & linear absorption; reflective mask, vacuum & strong & \cite{Itani2013}\\
Electron-beam lithography & sub-10\,nm & electron--resist energy loss & weak (de Broglie) & \cite{Juodkazis2026stitch}\\
Multiphoton polymerisation (2PP/TPL) & $\sim$100--120\,nm & $k$-photon absorption above polymerisation threshold & nonlinear, sub-$\lambda$ & \cite{Kawata2001,Gattass2008,Davis1996,Glezer1996,Samsonas2023}\\
STED-inspired DLW & $\sim$50--55\,nm & depletion of excited state, $I_{\rm sat}$ & nonlinear & \cite{Hell1994,Fischer2013}\\
Microsphere-assisted fs structuring & $\sim$38\,nm ($\lambda/27$) & photonic-nanojet concentration; thermal-stress nanogrooves & moderate & \cite{Yin2025,Lin2025}\\
\blue{Dual-beam far-field patterning (Si)} & \blue{$\sim$10--12\,nm ($\lambda/66$)} & \blue{orthogonal-polarisation double-beam overlap; thermal-stress localisation} & \blue{moderate} & \blue{\cite{Lin2020NL}}\\
\blue{Cavity-seeded evanescent structuring (Bessel, glass)} & \blue{$\sim$7\,nm ($<\lambda/100$)} & \blue{far-field-induced cavities seed evanescent near-fields; local thermal ablation} & \blue{none (near-field)} & \blue{\cite{Zhang2025UFS}}\\
Deterministic threshold ablation & tens of nm & above-threshold MPI/avalanche on Gaussian tip\blue{; contour mapping at strict threshold~\cite{GarciaLechuga2020}} & weak & \cite{Liu1997,Joglekar2004}\\
Quantized interference structuring & $\sim\lambda/2n$ & threshold + standing-wave interference & moderate & \cite{Kumar2014}\\
\blue{Controlled-avalanche ripple$\,\to\,$single-groove ablation (sapphire)} & \blue{$\sim$50--70\,nm ($\lambda/6$)} & \blue{near-threshold avalanche; sub-surface plasma sphere-to-plane growth; ripple pattern collapses to one groove} & \blue{moderate ($\Lambda\!\approx\!\lambda/2n$ seed)} & \blue{\cite{Buividas2013}}\\
Far-field-induced near-field breakdown (O-FIB, ``nano-pen'') & $\sim$20\,nm & boundary-condition $E$-field enhancement (normal $D$ continuity) & none (near-field) & \cite{Li2020}\\
Bessel nano-seed merging / super-stealth dicing (``nano-knife'') & $\sim$10\,nm, AR $\sim10^3$ & near-field merging of ionised seeds along filament & none & \cite{Li2024,Ganguly2024}\\
ENZ / Die-Met dipole localisation & tens\,$\rightarrow$\,single nm & induced dipole, $I\propto r^{-6}$, $\varepsilon\!\to\!0$ & none & \cite{Gamaly2018,Juodkazis2026stitch}\\
\blue{Nanocavitation / void-seeded scattering} & \blue{$\sim$10\,nm} & \blue{pressure-gradient cavitation, rarefaction and fast quenching; void sustains field confinement} & \blue{none} & \blue{\cite{Bhuyan2017,Zhang2025UFS,Juodkazis2006}}\\
Tunnelling-ionisation defect inscription & atomic / quantum & polarisation-directed bond breaking & none & \cite{Wang2024,Castelletto2018}\\
Laser-seeded microwave volumetric deposition (this work) & tens\,$\rightarrow$\,single nm & laser defines lossy nanovolume; cavity microwave Joule-heats it; thermal runaway & none (absorber-defined) & ---\\
\end{tblr}
\end{table*}

\subsection{Far-field nonlinear deposition}

The pioneering demonstrations of sub-diffraction writing exploited nonlinear absorption. Two-photon polymerisation reached \SI{120}{nm} features below the
diffraction limit already in 2001~\cite{Kawata2001}, building on the discovery that tightly focused femtosecond pulses write permanent refractive-index changes and store data inside transparent solids~\cite{Davis1996,Glezer1996}. The mechanisms and parameter space of femtosecond micromachining in transparent materials were comprehensively reviewed 
~\cite{Gattass2008}.
Borrowing the stimulated-emission-depletion (STED) concept from super-resolution microscopy~\cite{Hell1994}, STED-inspired lithography pushed polymer feature sizes to $\sim\!\SI{55}{nm}$~\cite{Fischer2013}, the resolution improving as
$(1+I_{\max}/I_{\rm sat})^{-1/2}$. These methods remain nonlinear but wavelength-referenced and, in the STED case, demanding in dual-beam overlap and
resist chemistry; the damage/polymerisation threshold itself depends on wavelength and pulse duration, as quantified for 3D nanopolymerisation~\cite{Samsonas2023}.

\subsection{Alternative routes: cavitation, scattering centres and the material response}\label{sec:alt}

A balanced reading of the recent literature shows that the transient-absorber picture of this review is one member of a family of mechanisms, and we state the alternatives explicitly.

\textbf{Threshold without nonlinearity.} As noted in Sec.~\ref{sec:threshold}, Garcia-Lechuga \emph{et al.}~\cite{GarciaLechuga2020} demonstrated that single-shot ablation contours map the beam profile at a strict threshold intensity independent of the absorption order. The nonlinearity therefore sharpens the \emph{dose}, but the edge definition of an ablation crater can be a purely threshold-contour property.
Eq.~(\ref{eq:threshold}) shows that this holds for any monotonic dose threshold: at a given $I_0/I_{\rm th}$ the contour radius does not depend on $k$. The nonlinearity sharpens the dose gradient at the contour and confines absorption to the focal volume, but it does not by itself reduce the modified radius below the intensity-threshold contour.

\textbf{Nanocavitation and rarefaction.} A void or rarefied (low-density) region formed by pressure-gradient cavitation under strong confinement and fast quenching is itself a low-permittivity scattering centre, with no requirement that the carrier density ever approaches the critical value. The ``slow microexplosion'' in bulk silica~\cite{Bhuyan2017} and the cavity-seeded evanescent-field structuring of Zhang \emph{et al.}~\cite{Zhang2025UFS} show that, once such a centre exists, it sustains scattering and near-field confinement during subsequent writing. The same electromagnetic role that the ENZ/Die-Met seed plays in Sec.~\ref{sec:dipole}, reached by a mechanical rather than electronic route. Because quenching can be arbitrarily fast under confinement, and mechanical processes such as interfacial sliding have interatomic action range, this pathway has no obvious lower length-scale bound. The two routes are complementary: the internal-field enhancement of Eq.~(\ref{eq:internal3D}) depends only on the permittivity contrast, however generated. The ENZ transit is thus a sufficient, self-sharpening route to a low-$\varepsilon$ seed within a single pulse, but it is \emph{not a sine qua non} for near-field-localised writing.

\textbf{Surface instabilities and metals.} Sub-surface cavitation and hydrodynamic instabilities, observed for instance in ripple formation, provide local field and energy-deposition enhancement even in metals, where free carriers pre-exist and no dielectric-to-metal conversion is needed. Polarisation-coupled scattering at such corrugations is the accepted origin of LIPSS. Our framework addresses the complementary case of initially transparent media, where the transient conversion is the self-sharpening step.

\textbf{Transformation mechanisms.} Field localisation fixes where energy arrives, i.e., the conversion into a permanent feature proceeds by melting and resolidification, densification/rarefaction, cavitation, defect accumulation or, in the extreme normal-field case, electrostatic (Coulomb-explosion-type) ion removal~\cite{Gamaly2023}. The latter requires the surface charge to survive longer than the ion response time 
before screening by conduction electrons neutralises the surface charge and suppresses Coulomb explosion, leaving thermal/hydrodynamic routes to dominate~\cite{Gamaly2018}. Experimental verification of the layer-by-layer electrostatic regime remains outstanding (Sec.~\ref{sec:polar}).

\subsection{The atomic and sub-atomic limits}

Beyond $\sim\!\SI{10}{nm}$, deterministic single-defect inscription becomes the goal. \blue{Polarisation-sensitive strong-field (tunnelling) ionisation}~\cite{Wang2024,Castelletto2018,Castelletto2021,Choi2016} \blue{can in principle direct bond breaking and create individual optically/magnetically active defects. The role of the tunnelling channel is that, for Keldysh parameter $\gamma\lesssim1$~\cite{Keldysh1965}, the ionisation rate follows the instantaneous field direction, so bonds aligned with $\bm E$ are ionised preferentially. Directionality of multiphoton absorption, which samples the cycle-averaged intensity, largely lacks} in atomically thin h-BN the uncertainty was found to be set by the thermal motion of atoms rather than by the gradient of the optical envelope, and the emission turned from stochastic/polychromatic to deterministic/monochromatic~\cite{Wang2024}. At
relativistic intensities ($\gtrsim\!10^{18}\,\mathrm{W\,cm^{-2}}$) two distinct
frontiers open. The first is structural: confined micro-explosion synthesizes high-pressure phases and superdense materials that are quenched to ambient
conditions: tetragonal Si polymorphs and superdense
aluminium~\cite{Rapp2024,Rapp2015,Vailionis2011}, a route to new phases. 
The second is genuinely nuclear and defines the
ultimate single-atom limit of material processing: transmutation, the change of one element into another. Few-cycle relativistic pulses accelerate
deuterons that drive the $^{2}\mathrm{H}(^{2}\mathrm{H},n)^{3}\mathrm{He}$ fusion
reaction, and compact Extreme Light Infrastructure (ELI)-class sources have demonstrated laser-driven D-D fusion with fast-neutron yields already at the $10^{5}$\,n/s level and a credible roadmap
to $\sim\!10^{10}$\,n/s at kHz repetition rate using few-cycle pulses~\cite{Osvay2024neutron,Osvay2024roadmap}. Whereas every method above localises energy to a small volume, transmutation localises identity to a single nucleus, the true atomic endpoint of the resolution hierarchy developed in this review. The near-field and absorber-defined mechanisms, to which we now turn, occupy the crucial middle ground between these limits and the diffraction-bound far-field methods.

\section{The unifying mechanism: dipole / ENZ near-field localisation} \label{sec:dipole}
The concept that unifies the wavelength-independent entries of Table~\ref{tab:survey} is that a sub-wavelength ionised region of reduced permittivity behaves as an induced dipole whose near-field is strongly localised. \blue{Two terms recur below and are not synonyms: \emph{Die-Met} denotes the entire transient conversion of the excited solid from dielectric ($\varepsilon'_{\rm in}>1$) to metallic ($\varepsilon'_{\rm in}<0$) response as the carrier density grows~\cite{Gamaly2018}. The \emph{ENZ} denotes the single waypoint $\varepsilon'_{\rm in}\!\to\!0$ on that trajectory, reached near (not at) the critical density once the background permittivity, collision frequency and effective mass are accounted for. Each carrier-density milestone on the trajectory has distinct physical properties, catalogued below and, as a time sequence, at the end of this section.}

\begin{figure*}[tb]
\centering\includegraphics[width=1\linewidth]{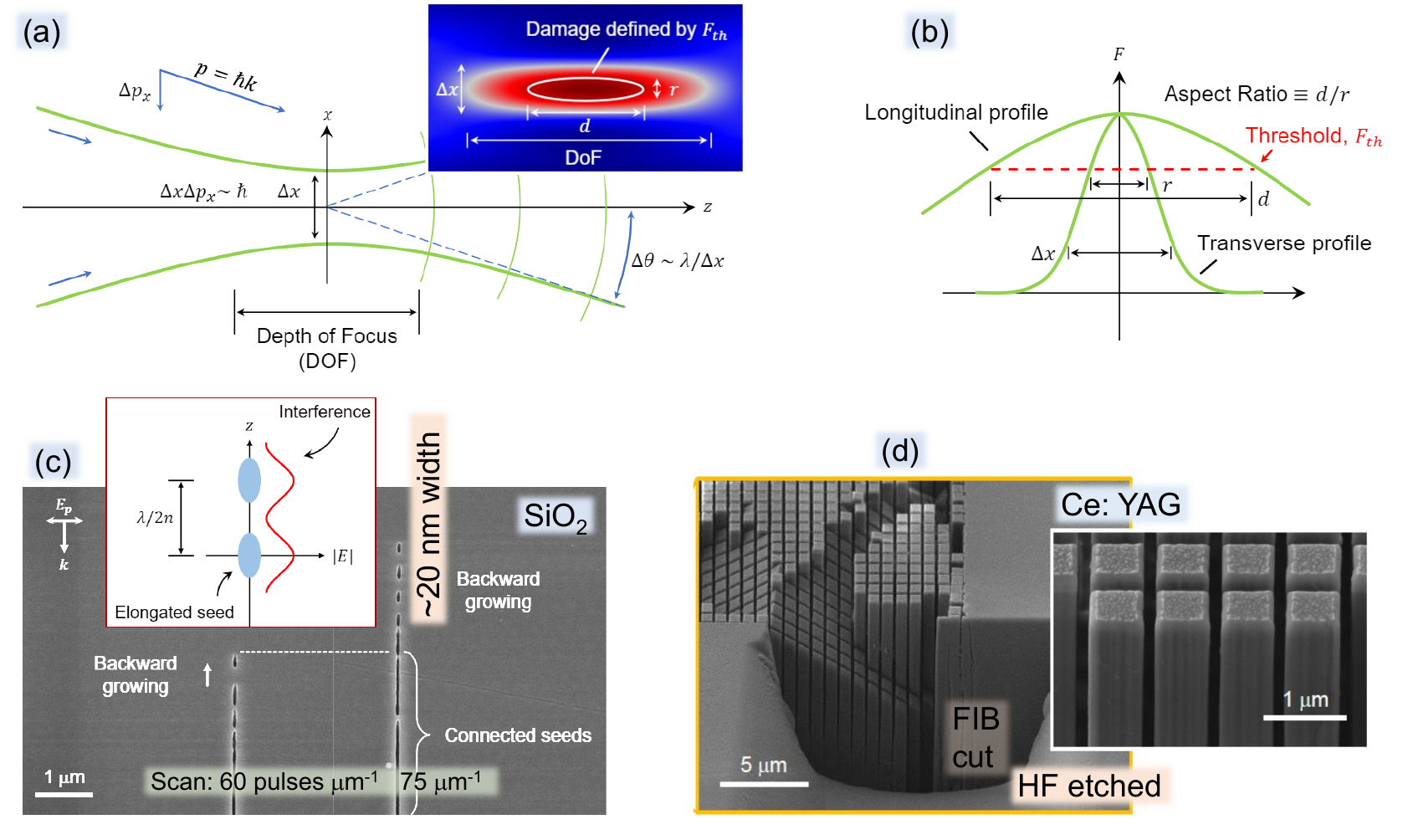}
\caption{Breaking the limit in aspect ratio of 3D modification (adapted from ref.~\cite{
Li2024}). (a) Equivalence of the Heisenberg's uncertainty principle and the waist-divergence relation for the Gaussian pulse (standard notations:
$p, k, \hbar$). The aspect ratio DoF/waist of the laser focus is limited; the same is valid for a Bessel-beam
axially long pulse, limited at a higher ratio. (b) The threshold effect: the damage/modification region
defined by a threshold fluence $F_{th}$ reduces the transverse diameter $r < \Delta x$ and the depth $d < DoF$ of the laser-modified volume simultaneously. Therefore the final aspect ratio of laser-modified
volume is limited by the shape of laser focus $d/r \sim DoF/\Delta x$. (c) Side-view SEM image of laser cut (long focus was scanned with overlap in out-of-page direction). Nano-seed structural damage sites of silica glass are merging into a nano-line via near-field enhancement, which is still localised
at nanoscale. (d) Breaking the limit: merged nano-seeds into nanoline cuts crystal at the record aspect ratio $\sim 10^4$~\cite{Li2024}.} \label{f-Heis}
\end{figure*}

\subsection{The induced dipole and the $I\propto r^{-6}$ law}

Consider the ionised seed as a sphere of radius $a\ll\lambda$ and (Drude) permittivity
\begin{equation}
\varepsilon_{\rm in}(\omega)=\varepsilon_b-\frac{\omega_p^2}{\omega^2+i\omega\nu},
\qquad
\omega_p^2=\frac{n_e e^2}{\varepsilon_0 m^\ast},
\label{eq:drude}
\end{equation}
embedded in a host of permittivity $\varepsilon_h$. Here $\omega$ is the optical angular frequency, $\varepsilon_b$ the bound-electron (background) permittivity, $\varepsilon_0$ the vacuum permittivity, $\omega_p$ the plasma frequency, $e$ the
elementary charge, $n_e$ the laser-generated carrier density, $\nu$ the collision frequency and $m^\ast$ the carrier effective mass. In the quasi-static limit the sphere acquires a dipole moment $\bm p=4\pi\varepsilon_0\varepsilon_h a^3\,\beta\,\bm E_{\rm loc}$ with the Clausius-Mossotti factor~\cite{Jackson1999,BohrenHuffman1983}
\begin{equation}
\beta=\frac{\varepsilon_{\rm in}-\varepsilon_h}{\varepsilon_{\rm in}+2\varepsilon_h}.
\label{eq:CM}
\end{equation}
It is essential to follow the real part of the permittivity as the laser-driven carrier density $n_e$ grows during the pulse. From Eq.~(\ref{eq:drude}), $\varepsilon'_{\rm in}\simeq\varepsilon_b-\omega_p^2/(\omega^2+\nu^2)$ falls monotonically from its bulk value. It first crosses below unity into the epsilon-near-zero (ENZ) window $0<\varepsilon'_{\rm in}<1$. The consequences for energy deposition are real but geometry-dependent and should not be reduced to a single velocity argument: near ENZ the wave impedance, the internal field distribution, the absorption length and the (dispersive, loss-corrected) energy density are all strongly modified, and ENZ media are known to enhance local fields and light-matter interaction under appropriate loss, angle and boundary conditions~\cite{Khurgin2021,Xie2025,Gamaly2018}. We do not invoke a simple $v_g\!\propto\!\sqrt{\varepsilon'_{\rm in}}$ slow-light relation (valid only for an idealised lossless, non-dispersive medium), nor a universal interface reflection minimum at a bulk dielectric/ENZ interface. The reflectance can be high unless loss, finite thickness or impedance matching intervene. The endpoint $\varepsilon'_{\rm in}=0$ is doubly significant: it is the perfect-absorber, dielectric-to-metal (``Die-Met'') condition at which the
deposited-energy density peaks~\cite{Gamaly2018}, and it is reached near the critical density. For the Drude form, $\varepsilon'_{\rm in}=0$ requires $\omega_p^2=\varepsilon_b(\omega^2+\nu^2)$, i.e.\
$n_{\rm ENZ}=\varepsilon_b\,(1+\nu^2/\omega^2)\,(m^\ast/m_e)\,n_c$ with the bare
vacuum-plasma value $n_c=\varepsilon_0 m_e\omega^2/e^2$. At $\lambda=\SI{1030}{nm}$
($\omega=\SI{1.83e15}{rad/s}$) the bare value is $n_c\approx\SI{1.05e21}{cm^{-3}}$
(for $m^\ast=m_e$, $\varepsilon_b=1$); in a real dielectric ($\varepsilon_b\!\sim\!2$--$12$) the ENZ density is correspondingly higher and depends on $m^\ast$, $\nu$ and band filling. Either way it is comparable to the $\sim\!10^{21}\,\mathrm{cm^{-3}}$ carrier density of the laser-formed seed. The useful operating window is therefore the controlled descent $\varepsilon'_{\rm in}\!:\,1\!\to\!0$, terminating at the ENZ/breakdown point. Past it ($\varepsilon'_{\rm in}<0$), the seed turns metallic and reflective, and the Fr\"ohlich resonance $\varepsilon'_{\rm in}=-2\varepsilon_h$ lying in that regime makes the
Clausius-Mossotti factor Eq.~(\ref{eq:CM}) diverge (limited only by losses) as
$n_e$ sweeps transiently through it. Because the internal field of a sphere is
$\bm E_{\rm int}=[3\varepsilon_h/(\varepsilon_{\rm in}+2\varepsilon_h)]\bm
E_{\rm loc}$, approaching ENZ concentrates the field inside the seed (by a
factor $3\varepsilon_h/2\varepsilon_h=1.5$ in the $\varepsilon_{\rm in}\!\to\!0$
limit) and suppresses the usual $\pi$ reflection phase so that incoming energy
couples efficiently. The
dipole near-field outside the seed~\cite{Jackson1999},
\begin{equation}
\bm E_{\rm nf}(\bm r)=\frac{1}{4\pi\varepsilon_0\varepsilon_h}
\frac{3(\bm p\cdot\hat{\bm r})\hat{\bm r}-\bm p}{r^3},
\label{eq:nearfield}
\end{equation}
yields an intensity
\begin{equation}
I(r)\propto|\bm E_{\rm nf}|^2\propto r^{-6},
\label{eq:r6}
\end{equation}
the quasi-static dipole field decays as $E\propto r^{-3}$, so intensity falls as $r^{-6}$. The prefactor depends on the seed polarisability, size and loss, so this is a localisation scaling (the dose width is a fixed fraction of the seed size
$a$, a length tied to $a$, not to $\lambda$), not a free-standing resolution law~\cite{Juodkazis2026stitch}. We stress that the ENZ condition ($\varepsilon'_{\rm in}\!=\!0$, internal field only $1.5E_0$), optical breakdown (the
critical density), the Fr\"ohlich dipole resonance
($\varepsilon'_{\rm in}\!=\!-2\varepsilon_h$, a lossy/metallic state with large but loss-limited enhancement) and the perfect-absorber condition are distinct states at different carrier densities, not one continuous favourable regime. \blue{Moreover, as developed in Sec.~\ref{sec:alt}, the electromagnetic role of the low-$\varepsilon$ seed can equally be played by a cavitation void or rarefied region generated mechanically: the ENZ transit is a self-sharpening route to the seed, not a prerequisite for the near-field localisation itself.} Combining Eq.~(\ref{eq:r6}) with the threshold relation Eq.~(\ref{eq:threshold}), a
$w_0\simeq\SI{0.5}{\micro\meter}$ focus operated near threshold produces $a\sim\SI{10}{}$-$\SI{30}{nm}$, in agreement with the demonstrated nano-pen and
nano-knife features~\cite{Li2020,Li2024}.

\blue{\textbf{Temporal sequence during the pulse and how to probe it.} For a near-threshold pulse of duration $\tau_p\sim200$\,fs the seed traverses these states in a definite order: (i) during the leading edge, multiphoton/tunnelling ionisation raises $n_e$ and $\varepsilon'_{\rm in}$ descends from $\varepsilon_b$ through unity via weak perturbation, little deposition; (ii) around the pulse peak the ENZ window is crossed, impedance mismatch collapses, the internal field rises to $1.5E_0$ [Eq.~(\ref{eq:internal3D})] and the volumetric deposition rate $\propto\varepsilon''_{\rm in}|\bm E_{\rm int}|^2$ peaks. We expect the dominant share of the absorbed energy to be deposited in this stage. Then, (iii) if ionisation continues, $\varepsilon'_{\rm in}$ sweeps through the Fr\"ohlich value $-2\varepsilon_h$, giving a transient, loss-limited spike of the external near-field. Followed by (iv) deeper into the metallic branch the seed reflects, deposition migrates to a skin layer and to the external near-field lobes, and self-limits. The sequence is experimentally accessible: time-resolved reflectivity/transmission pump-probe traces the $\varepsilon'(t)$ descent, THz emission tracks the transient currents in the focal volume~\cite{Huang2024}, and four-polarisation near-field imaging resolves the anisotropy of the final deposition~\cite{Ryu2022,Honda2019}. A measured crossing time of the reflectivity minimum (ENZ) relative to the pulse peak would directly test stage (ii), the central claim of this section.}

\subsection{2D writing on surface: the nano-pen}

The boundary conditions make this concrete in 2D. The continuity of the normal component of the displacement field, $\varepsilon_1 E_{1n}=\varepsilon_2 E_{2n}$~\cite{Jackson1999},
enhances the $E$-field on the low-permittivity side of an interface. This applies to lossless dielectrics; at a conducting interface the relevant continuity is that of the normal current density. Charge conservation across the
boundary gives, for harmonic fields, continuity of the normal total (conduction\,$+$\,displacement) current,
\begin{equation}
\big(\sigma_1-i\omega\varepsilon_0\varepsilon_1\big)E_{1n}
=\big(\sigma_2-i\omega\varepsilon_0\varepsilon_2\big)E_{2n},
\label{eq:condBC}
\end{equation}
i.e.\ continuity of the normal component of the complex permittivity $\tilde\varepsilon=\varepsilon+i\sigma/(\varepsilon_0\omega)$~\cite{Jackson1999}. In the quasi-static conduction limit Eq.~(\ref{eq:condBC}) reduces to $\sigma_1E_{1n}=\sigma_2E_{2n}$: the normal field is enhanced on the low-conductivity side, the conduction-current twin of the dielectric law, with the loss tangent $\sigma/(\omega\varepsilon_0\varepsilon)$ deciding which term dominates.

Silicon and copper bracket the two regimes. At a \SI{1030}{nm} optical frequency ($\omega\varepsilon_0\approx1.6\times10^{4}\,\mathrm{S\,m^{-1}}$), copper
($\sigma_{\rm Cu}\approx6\times10^{7}\,\mathrm{S\,m^{-1}}$) has
$\sigma/(\omega\varepsilon_0)\approx4\times10^{3}$ (and $\approx4\times10^{8}$ at \SI{2.45}{GHz}): it is metallic at every relevant frequency, the field is expelled into a nanometric skin depth and reflected; the enhancement lives entirely in the adjacent dielectric. Intrinsic silicon ($\sigma_{\rm Si}\sim10^{-3}\,\mathrm{S\,m^{-1}}$, $\varepsilon_{\rm Si}\approx12$) has $\sigma/(\omega\varepsilon_0)\sim10^{-7}$ optically and behaves as a near-ideal dielectric, so the field penetrates and the dielectric law $\varepsilon_1E_{1n}=\varepsilon_2E_{2n}$ applies. This contrast is precisely why silicon is an especially favorable direct-write medium: it starts on the dielectric branch, and the pulse converts the focal nanovolume into a conducting ENZ/Die-Met seed. $\sigma$ rising by $\sim\!10^{10}$ as
$n_e\!\to\!10^{21}\,\mathrm{cm^{-3}}$, so the seed flips locally onto the metallic branch where it was written. The key requirement is a large transient permittivity/conductivity change during the pulse, which yields a self-sharpening seed. Metals (skin-depth absorption, plasmonic field concentration, electron-phonon nonequilibrium) and passive dielectrics (high-NA nonlinear absorption) also localise energy, but by different, less self-sharpening routes. The self-induced near-field localisation emphasised here is strongest in media that traverse $\sigma\!\sim\!\omega\varepsilon_0$ within the pulse.

In far-field-induced near-field breakdown (O-FIB), a diffraction-limited pulse first ablates a $\sim\!\SI{15}{nm}$ seed hole; at a reduced fluence ($\sim\!0.8\,F_{\rm th}$, i.e.\ $\sim\!75\%$ of threshold) only the nanoscale vicinity of that hole is field-enhanced, so a $\sim\!\SI{20}{nm}$ groove is written by steering the far-field beam~\cite{Li2020}. This ``nano-pen'' writes 2D etch-mask patterns in a dielectric nanofilm in ambient air, without vacuum or high-precision stages, and, decisively, its resolution is set by the interface enhancement, not by the wavelength. Crucially, the direction of writing is set by the laser polarisation, the subject of Sec.~\ref{sec:polar}.

\subsection{3D writing inside volume: the nano-knife}

Three-dimensional recording, direct laser writing of structures within the volume of a transparent solid, is where the near-field, wavelength-independent localisation pays off most. It first reached the true nanoscale with Bessel-beam ``super-stealth dicing'' 
~\cite{Li2024} [Fig.~\ref{f-Heis}(c,d)], and the same dipole/near-field mechanism carries over to an ordinary Gaussian (cylindrical) focus~\cite{Li2026}. The same physics as for 2D surface patterning extends into the bulk. Along a Bessel-like filament, nanoscale seed modifications generated at the intensity maximum reduce the local permittivity; incident light back-scattered by the seeds interferes with the incoming beam in a ``crawling'' positive-feedback loop that homogenises the longitudinal deposition and confines the transverse field, merging the seeds into a continuous nano-plane of $\sim\!\SI{10}{nm}$ width~\cite{Li2024} [Fig.~\ref{f-Heis}(c)]. A Gauss-law identity for the Bessel focus places the top $25\%$ of the intensity within a $\sim\!\SI{25}{nm}$-wide channel, the seed of the cut. This ``super-stealth dicing'' achieves a record aspect ratio length/width of order $10^3$ (with the longest filaments approaching $10^4$)~\cite{Li2024}, complementary to axicon-doublet ultra-high-aspect machining in silicon~\cite{Ganguly2024}\blue{, to the cavity-seeded $\lambda/100$-class trenching of Zhang \emph{et al.}~\cite{Zhang2025UFS}, and to nanohole-clad waveguides with depth-to-diameter ratios up to $5\times10^4$~\cite{Fan2026LAM}}. Here the 3D localisation is again governed by the self-generated low-permittivity seed, a dipole-like absorber, rather than by $\lambda$. Why the energy concentrates at the reduced-permittivity centre and how the polarisation steers it for an ordinary Gaussian focus as well as a Bessel filament is developed 
in Sec.~\ref{sec:3Dconc}.
\begin{figure*}[t]\centering
\centering\includegraphics[width=1\linewidth]{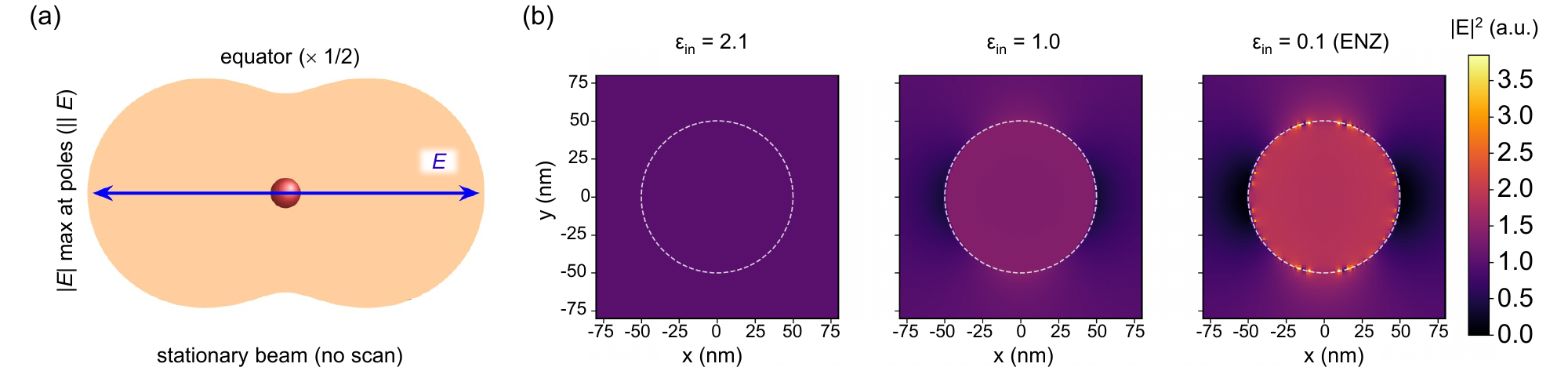}
\caption{Local field at a stationary focus. (a) The induced dipole's quasi-static near-field [Eq.~(\ref{eq:nearfield})] is twice as strong along $\bm E$ (poles) as across it (equator), a $4{:}1$ intensity ratio [Eq.~(\ref{eq:4to1})]. A defect or modification inscribed without scanning is expected to be elongated along the polarisation. (b) FDTD simulated $|E|^2$ field maps in the XY-plane containing the polarisation axis ($\bm{E} \parallel x$) for three sphere permittivities $\varepsilon_{\rm in}$ (SiO$_2$ host, $\varepsilon_h = 2.1$, $a = 50$\,nm, $\lambda = 1030$\,nm). As $\varepsilon_{\rm in}$ decreases toward zero (ENZ), the field concentrates progressively inside the sphere, consistent with [Eq.~(\ref{eq:internal3D})].}
\label{fig:inscribe}
\end{figure*}
\section{Polarisation control of dipole energy deposition}\label{sec:polar}

The seed oscillates as a dipole $\bm p\parallel\bm E$, so the deposited energy is polarisation-dependent. At a stationary focus the quasi-static near-field is
strongest along $\bm E$, orienting the inscribed defect even without scanning (Fig.~\ref{fig:inscribe}). During scanning, the writing of an extended groove (on surface) is governed by the
quasi-static near-field and the electrostatic boundary conditions at the groove wall, which place the strongest deposition transverse to $\bm E$ (Fig.~\ref{fig:polar}). Both regimes are
near-field; the far-field dipole radiation pattern 
is not the deposition mechanism. We treat the cases in turn.

\subsection{Stationary focus: local-field anisotropy and oriented defects}

With the beam held fixed, the relevant field is the quasi-static near-field of the induced dipole, Eq.~(\ref{eq:nearfield}). Its magnitude along the polarisation axis (the poles, $\theta=0$) and across it (the equator, $\theta=\pi/2$), at the same radius, is
\begin{equation}
|\bm E_{\rm nf}|_{\rm pole}=\frac{2p}{4\pi\varepsilon_0\varepsilon_h r^3},\qquad
|\bm E_{\rm nf}|_{\rm eq}=\frac{p}{4\pi\varepsilon_0\varepsilon_h r^3},
\label{eq:poleq}
\end{equation}
so the field is twice as large along $\bm E$ as across it and the deposited dose ($\propto|E|^2$) obeys
\begin{equation}
\frac{I_{\rm pole}}{I_{\rm eq}}=4 .
\label{eq:4to1}
\end{equation}
This four-to-one anisotropy is exact for a point dipole~\cite{Jackson1999} and is active during the inscription of a single modification: a colour centre, dangling bond, or amorphised voxel even with no scanning (Fig.~\ref{fig:inscribe}). The energy deposition, and hence \blue{the envelope within which the defect is generated,} is elongated along $\bm E$\blue{. Two distinct effects should not be conflated here: (i) the deposition (damage-probability) envelope around the seed is elongated along $\bm E$ by the 4:1 near-field anisotropy of Eq.~(\ref{eq:4to1}), a statement about where a defect is likely to form, not about the shape of a point defect and (ii) the microscopic orientation of the created defect can follow the field direction through bond-aligned strong-field ionisation (Sec.~\ref{sec:survey}). The h-BN experiments provide evidence for deterministic, polarisation-correlated defect creation at the quantum-confinement scale~\cite{Wang2024}, while} the deterministic polarisation control \blue{of ablation morphology was demonstrated on metal surfaces}~\cite{Iwase2009}. Once the modification extends into a groove, the complementary radiation/boundary-condition picture below takes over.

\subsection{Near-field, not radiation, governs the local deposition}

A clarification is essential. The \emph{far-field} radiation of an oscillating dipole follows the toroidal pattern
\begin{equation}
I(\theta)\propto\sin^2\theta ,
\label{eq:sin2}
\end{equation}
with $\theta$ measured from $\bm E$, maximal perpendicular to $\bm E$ and vanishing along it. At the nanometric distances that matter for writing ($r\ll\lambda$), however, the radiative $1/r$ term is negligible: the field is dominated by the quasi-static near-field [$\propto1/r^3$, Eq.~(\ref{eq:nearfield})] and by the electrostatic boundary conditions at the seed/groove interface. The local deposition is therefore set by the near-field and boundary conditions, consistent with the seed being a sub-wavelength, non-radiating absorber, not by the radiation torus. Equation~(\ref{eq:sin2}) is retained only to note that the far-field
anisotropy and the near-field boundary enhancement share the same symmetry: both
place the strongest transverse response across $\bm E$.

\subsection{Boundary-condition enhancement inside the groove}

\begin{figure*}[t]\centering
\includegraphics[width=1\linewidth]{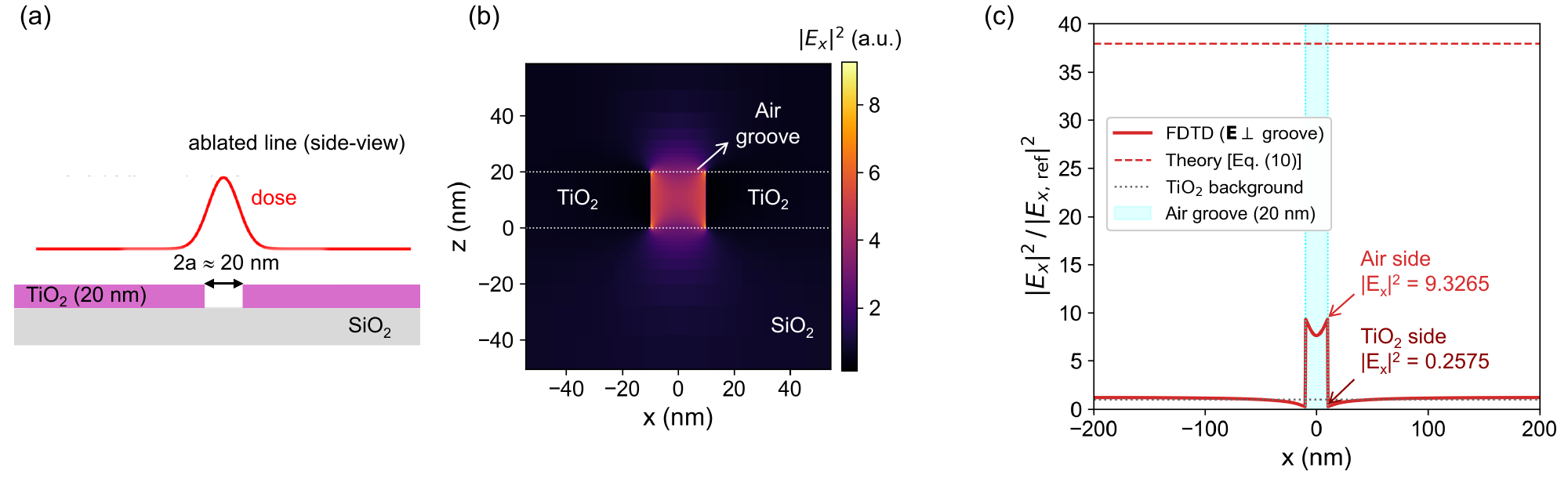}
\caption{Polarisation and the written groove. (a) Profile of the line written by the nano-pen: a groove of width $\approx\!2a\approx\SI{20}{nm}$ in a \SI{20}{nm} titania film on silica, its width set by near-field boundary-condition enhancement and the seed size, not by $\lambda$~\cite{Li2020}. (b) FDTD simulated $|E_x|^2$ field map ($\bm{E}\perp$ groove, $\varepsilon_{\rm film}=6.16$, groove $20\times20$\,nm, $\lambda=1030$\,nm) in the $xz$ plane: the normal field is strongly enhanced inside the low-permittivity air groove, with hotspots at the groove corners. (c) Horizontal profile of $|E_x|^2$ across the groove wall at mid-height ($z=10$\,nm), normalised to the TiO$_2$ background. The measured peak intensity ratio across the groove wall is 36.22 (air side $|E_x|^2=9.3265$, TiO$_2$ side
$|E_x|^2=0.2575$), consistent with the electrostatic prediction $(\varepsilon_{\rm film}/\varepsilon_{\rm gr})^2=38.0$ [Eq.~(\ref{eq:Dnorm})]; the shortfall reflects the finite groove geometry and off-corner measurement position.}
\label{fig:polar}
\end{figure*}

The electrostatic boundary conditions point the same way and supply the quantitative enhancement. At the wall of an ablated groove, the normal component of $\bm D$ is continuous, $\varepsilon_{\rm film}E^{\perp}_{\rm film}=\varepsilon_{\rm gr}E^{\perp}_{\rm gr}$,
so a field \emph{normal} to the wall is enhanced inside the low-permittivity groove by~\cite{Li2020}:
\begin{equation}
\frac{E^{\perp}_{\rm gr}}{E^{\perp}_{\rm film}}
=\frac{\varepsilon_{\rm film}}{\varepsilon_{\rm gr}}.
\label{eq:Dnorm}
\end{equation}
For an air groove in a titania film ($\varepsilon_{\rm film}\approx6$, $n\approx2.5$) the internal field is enhanced $\sim\!6$-fold and the intensity $\sim\!\varepsilon_{\rm film}^2\approx40$-fold for the \emph{normal} component, an order-of-magnitude electrostatic estimate for an idealised planar gap.
\blue{For the nano-pen groove of Fig.~\ref{fig:polar}(a), FDTD gives a peak intensity ratio of 36 across the groove wall (Fig.~\ref{fig:polar}(b,c)).}
The actual enhancement at a finite-curvature, lossy, evolving groove wall will differ. A field \emph{tangential} to the wall is continuous and not enhanced. This normal-field boundary enhancement is the near-field mechanism we associate with the
nano-pen~\cite{Li2020}, placing the strongest deposition transverse to $\bm E$.

\subsection{The writing rule: polarisation perpendicular to the scan}

A groove runs along the scan direction, so its long walls are parallel to the scan.
To enhance the field inside the groove the polarisation must be \emph{normal}
to those walls, that is, perpendicular to the scan (Fig.~\ref{fig:polar}). This is exactly the demonstrated nano-pen: a $\sim\!\SI{20}{nm}$ groove ablated in a
\SI{20}{nm} titania film on silica, its width fixed by the normal-$\bm D$ enhancement of Eq.~(\ref{eq:Dnorm}) and the seed size $\approx\!2a$ rather than by the wavelength~\cite{Li2020} [Fig.~\ref{fig:polar}]. The enhanced normal field inside the groove deposits energy at the advancing wall and propagates the groove
along the scan. With the polarisation instead parallel to the scan, the field at the long walls is tangential, continuous and not enhanced, and no narrow groove forms. The polarisation thus selects the writing direction: the nano-pen draws its finest line
with $\bm E$ perpendicular to the scan, and the same rule governs the seed-merging of the nano-knife~\cite{Li2020,Li2024}. Rotating the polarisation by $90^\circ$ switches the narrow-groove writing off, the same polarisation control was used to suppress periodic ripples on ablated Ni~\cite{Iwase2009}. For completeness, we note that polarisation-independent nanostructuring is also possible when the deposition is dominated by thermal stress rather than by the near-field, as recently shown for microsphere-assisted femtosecond structuring~\cite{Yin2025}; the contrast between the two regimes underlines that the polarisation signature is a fingerprint of the near-field mechanism.

\blue{\textbf{From ripples to a single groove: tuning the energy deposition.} That the deposition can be continuously tuned between the two canonical surface morphologies, the periodic ripple (LIPSS) pattern and a single sub-diffraction groove, was demonstrated early on sapphire~\cite{Buividas2013}. Ripples with period $\Lambda\approx\lambda_0/(2n)$ form via plasma nano-planes nucleating in a sub-surface region (a sphere-to-plane growth of the ionised volume, seeded by accumulated defects) and are imprinted on the surface by ablation. As the exposure is brought to within $\sim\!10\%$ of the ablation threshold by reducing pulse energy or pulse-to-pulse overlap, the multi-groove pattern collapses to a single 50--70\,nm groove ($\sim\!\lambda/6$, one tenth of the $1.22\lambda/\mathrm{NA}=\SI{700}{nm}$ diffraction-limited spot at the NA$=0.7$, 400\,nm conditions used), written by a controlled avalanche breakdown. This is the same near-threshold logic as Eq.~(\ref{eq:threshold}), with the decisive practical advantage that the ripple-mediated route needs only $\sim\!10\%$ pulse-energy stability, whereas confining a bare Gaussian to a 50\,nm contour would demand $\sim\!1\%$. Polarisation enters here too: ripples parallel and perpendicular to the scan exhibit $\sim\!10\%$ different formation thresholds and slightly different periods, attributed to the anisotropic heat transport and defect accumulation along the written texture~\cite{Buividas2013}, an early quantitative instance of the scan/polarisation coupling formalised by the writing rule above, and a direct precedent for the seed-and-merge writing of the nano-pen and nano-knife~\cite{Li2020,Li2024}.}

\subsection{The normal-field (out-of-plane) case and the single-bond limit}

The third distinct geometry is polarisation \emph{normal} to the surface. At high-NA the non-paraxial focus already carries a longitudinal $E_z$, and a seed at an interface enhances the normal field by normal-$D$ continuity. Driving the field out of the plane opens the resonant-absorption channel and, in the extreme, a charged surface layer can undergo Coulomb explosion, removing material a monolayer
at a time according to the theory of ablation in a field normal to the surface via the layer-by-layer ion explosion in the sub-ion-period regime
~\cite{Gamaly2023}. \blue{
This is the route to atomic-precision surface removal that has no diffraction-limited analogue. The experimental verification of this mechanism is still expected.
}
The polarisation anisotropy persists all the way to the single bond. The quiver amplitude of a free electron in the optical field is $x_q=eE/(m_e\omega^2)$. At
$I=\SI{10}{TW/cm^2}$ ($E=\sqrt{2I/c\varepsilon_0}\approx\SI{8.7e9}{V/m}$) and
$\lambda=\SI{1030}{nm}$ ($\omega=\SI{1.83e15}{rad/s}$),
\begin{equation}
x_q=\frac{eE}{m_e\omega^2}
\approx 0.46~\mathrm{nm},
\label{eq:quiver}
\end{equation}
which exceeds the \SI{0.16}{nm} Si-O bond length in silica~\cite{Gamaly2018}. This
vacuum quiver amplitude is only a field-scale indicator, not a proof of
bond removal: the displacement of a bound electron is reduced by local-field
and screening factors and depends on the Keldysh parameter, collision time and band
structure, while the ions barely move on the pulse timescale ($x_q\!\propto\!\lambda^2$
makes the \SI{1030}{nm} field-scale $\sim\!1.7\times$ that at \SI{800}{nm}). What it
does show is that the optical field becomes comparable to the intra-bond field, so \blue{polarisation-sensitive strong-field (tunnelling) ionisation~\cite{Keldysh1965}} can preferentially break the bond aligned with $\bm E$ and inscribe an oriented defect, a plausible and experimentally indicated path to polarisation-controlled colour centres and single
dopants~\cite{Wang2024,Castelletto2018,Castelletto2021,Choi2016}. Demonstrating true single-bond determinism requires tunneling-rate/local-field modeling or atomistic simulation, not the quiver estimate alone.

\subsection{3D field concentration at a reduced-permittivity core}\label{sec:3Dconc}

\begin{figure}[t]\centering
\includegraphics[width=1\linewidth]{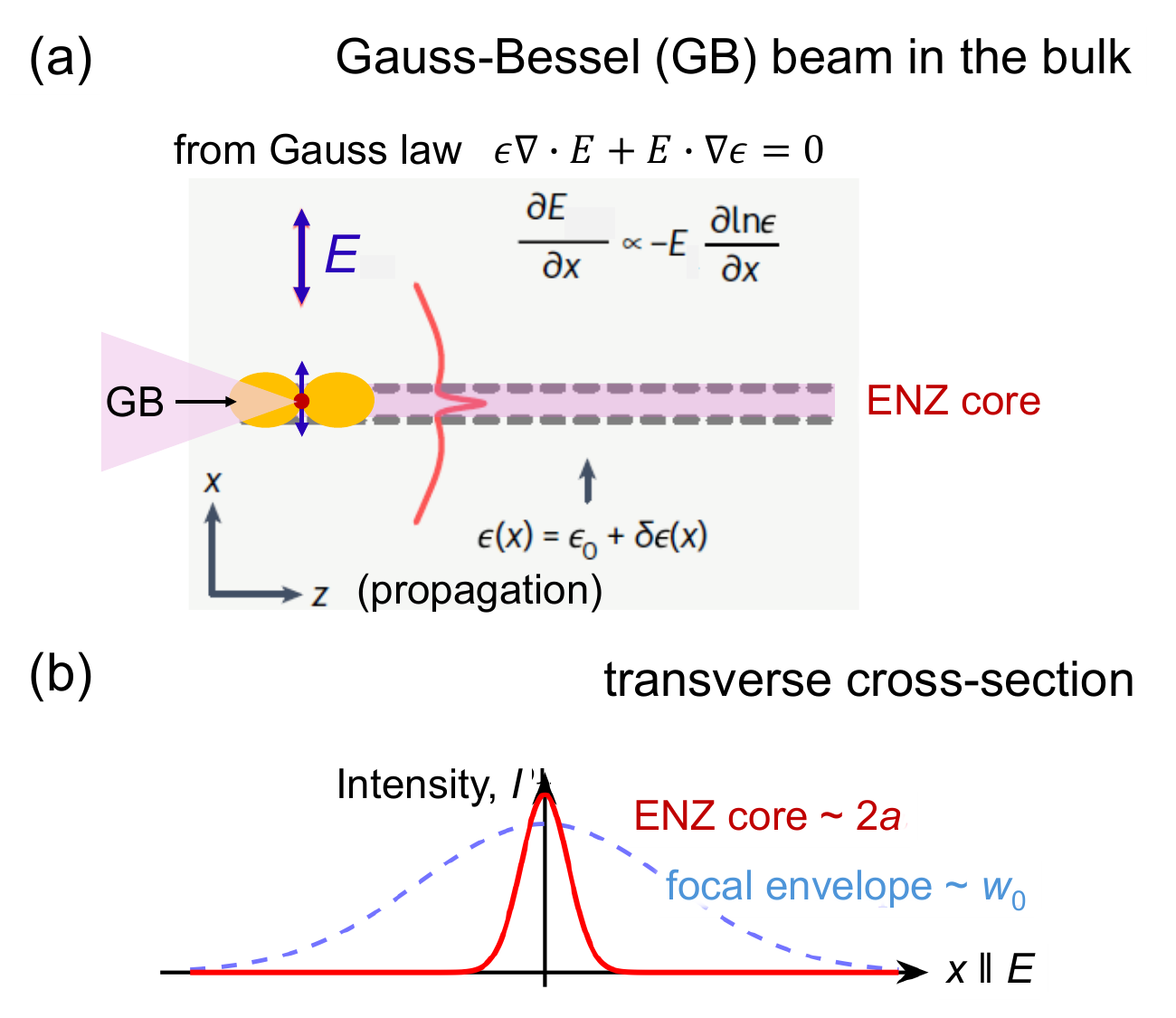}
\caption{Field concentration at a Gauss-Bessel (GB) focus inside the bulk (as in ref.~\cite{Li2024}); the same arguments are valid for a Gaussian beam. (a) The
diffraction-limited GB focus ($\sim\!w_0$) ionises a reduced-permittivity core; by
Eqs.~(\ref{eq:divE})-(\ref{eq:internal3D}) the field is pulled into that core and,
because only the component along $\bm E$ is amplified, the near-field is anisotropic
(lobes along the polarisation $\bm E$, transverse to the propagation $z$). (b)
Transverse intensity cut: the deposited dose (red, ``ENZ core'', FWHM $\sim\!2a$) is
far narrower than the optical focal envelope (blue dashed, $\sim\!w_0$) and is
steered by $\bm E$ for GB or Gaussian beam. This is a near-field effect localised to the core.}
\label{fig:gauss3d}
\end{figure}

The same dipolar localisation operates in the bulk, and a generic argument shows why the field pins itself to a region of depressed permittivity, whether that region is the core of a Bessel filament or the voxel of an ordinary Gaussian focus (Fig.~\ref{fig:gauss3d}). In a charge-free medium $\nabla\!\cdot\!\bm D=0$~\cite{Jackson1999}, so a field crossing a permittivity gradient acquires the divergence
\begin{equation}
\nabla\!\cdot\!\bm E=-\,\bm E\!\cdot\!\nabla\ln\varepsilon ,
\label{eq:divE}
\end{equation}
and the wave equation correspondingly carries the gradient term $\nabla(\bm E\!\cdot\!\nabla\ln\varepsilon)$~\cite{Li2024}; the component of $\bm E$ along $\nabla\varepsilon$ (normal to the seed boundary) is amplified where $\varepsilon$ is small.
For a field obliquely crossing a graded layer
$\varepsilon(x)$ this is exact, the in-plane component obeys
\begin{equation}
E_{\parallel\nabla\varepsilon}(x)=-\frac{\alpha_0}{\varepsilon(x)}\,G(x)
\;\xrightarrow[\;\varepsilon\to0\;]{}\;\infty ,
\label{eq:Esingular}
\end{equation}
with $\alpha_0=n_0\cos\theta$ ($n_0$ the host refractive index, $\theta$ the angle of
incidence) and $G(x)$ a slowly varying envelope that stays finite at the minimum. S, the field is pinned to the permittivity minimum and diverges at the ENZ point $\varepsilon\!\to\!0$~\cite{Li2024}. Equivalently, a sub-wavelength inclusion of reduced permittivity carries a uniform, enhanced
internal field~\cite{Jackson1999,BohrenHuffman1983},
\begin{equation}
\begin{aligned}
\bm E_{\rm int}&=\frac{3\varepsilon_h}{\varepsilon_{\rm in}+2\varepsilon_h}\,\bm E_0
&&(\text{sphere, Gaussian voxel}),\\
\bm E_{\rm int}^{\perp}&=\frac{2\varepsilon_h}{\varepsilon_{\rm in}+\varepsilon_h}\,\bm E_0
&&(\text{cylinder, Bessel filament}).
\end{aligned}
\label{eq:internal3D}
\end{equation}
rising to $1.5\,E_0$ and $2\,E_0$ respectively at ENZ, so the absorbed power
$\propto\varepsilon_{\rm in}''|\bm E_{\rm int}|^2$ self-focuses at the core. This is
the bulk counterpart of the surface boundary-condition enhancement of
Eq.~(\ref{eq:Dnorm}); identical physics whether the low-$\varepsilon$ region is a
filament (cylinder) or an ordinary focal voxel (sphere).

\begin{figure*}[tb]
\centering\includegraphics[width=.8\linewidth]{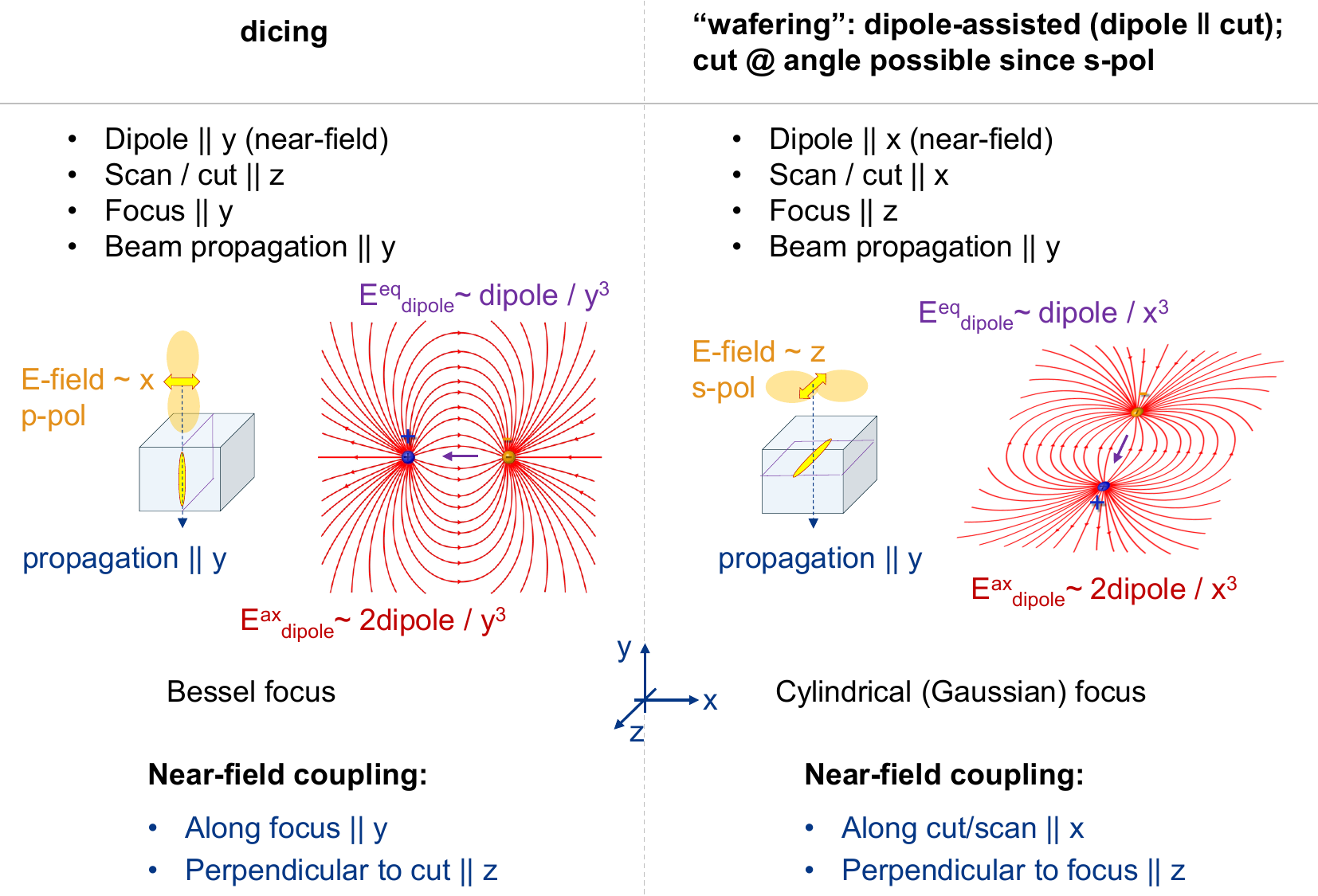}
\caption{Practical implementation of polarisation controlled cutting, (stealth-)dicing, ``wafering'' (tomo-slicing) by Gauss-Bessel (GB) beam~\cite{Li2024} and cylindrical focusing~\cite{Li2026}; see supplement in ref.~\cite{Li2026} for tomo-slicing conditions and mechanism of near-field energy deposition. Microwave energy deposition can be additional enhancement of this method by pre-heating during slicing or heating after laser inscription as hypothesized in this perspective review (Sec.~\ref{sec:paradigm}).
}\label{f-wafe}
\end{figure*}
The concentration is intrinsically polarisation-steered, exactly as for the
stationary near-field [Eq.~(\ref{eq:4to1})]: only the component along
$\bm E$ feels the $1/\varepsilon$ amplification of Eq.~(\ref{eq:Esingular}), so even for an ordinary Gaussian focus (Fig.~\ref{fig:gauss3d}) the core's near-field and the next increment of deposition is strongest at the poles along the polarisation, the inscribed voxel is correspondingly anisotropic, and rotating
$\bm E$ rotates its axis. The in-volume permittivity contrast is gentler than at an open surface (the host surrounds the seed on all sides rather than air on one), so the bulk anisotropy is weaker but of identical dipolar origin, the mechanism by which the cut plane and
width track the writing polarisation in super-stealth dicing~\cite{Li2024}.

Two cautions complete the picture, and both follow from this being a \emph{near-field} ENZ effect. First, the $1/\varepsilon$ amplification
[Eq.~(\ref{eq:Esingular})] and the $r^{-6}$ localisation [Eq.~(\ref{eq:r6})] are
quasi-static: they live within $\sim\!a$ of the seed and do not radiate, they fix where energy is deposited, not how light propagates in the far field. Second, the responsible region has reduced permittivity, i.e., a negative index change, which is anti-guiding and cannot support a bound propagating mode. This mechanism, therefore, does not by itself create a
propagating waveguide. A guiding (type-I) waveguide requires the opposite, a permanent positive index change from post-pulse densification or stress, a distinct material outcome from the transient ENZ energy deposition analysed here.

\section{Real-life engineering: direct write that already works} \label{sec:engineering}

\begin{figure*}
\centering\includegraphics[width=0.9\linewidth]{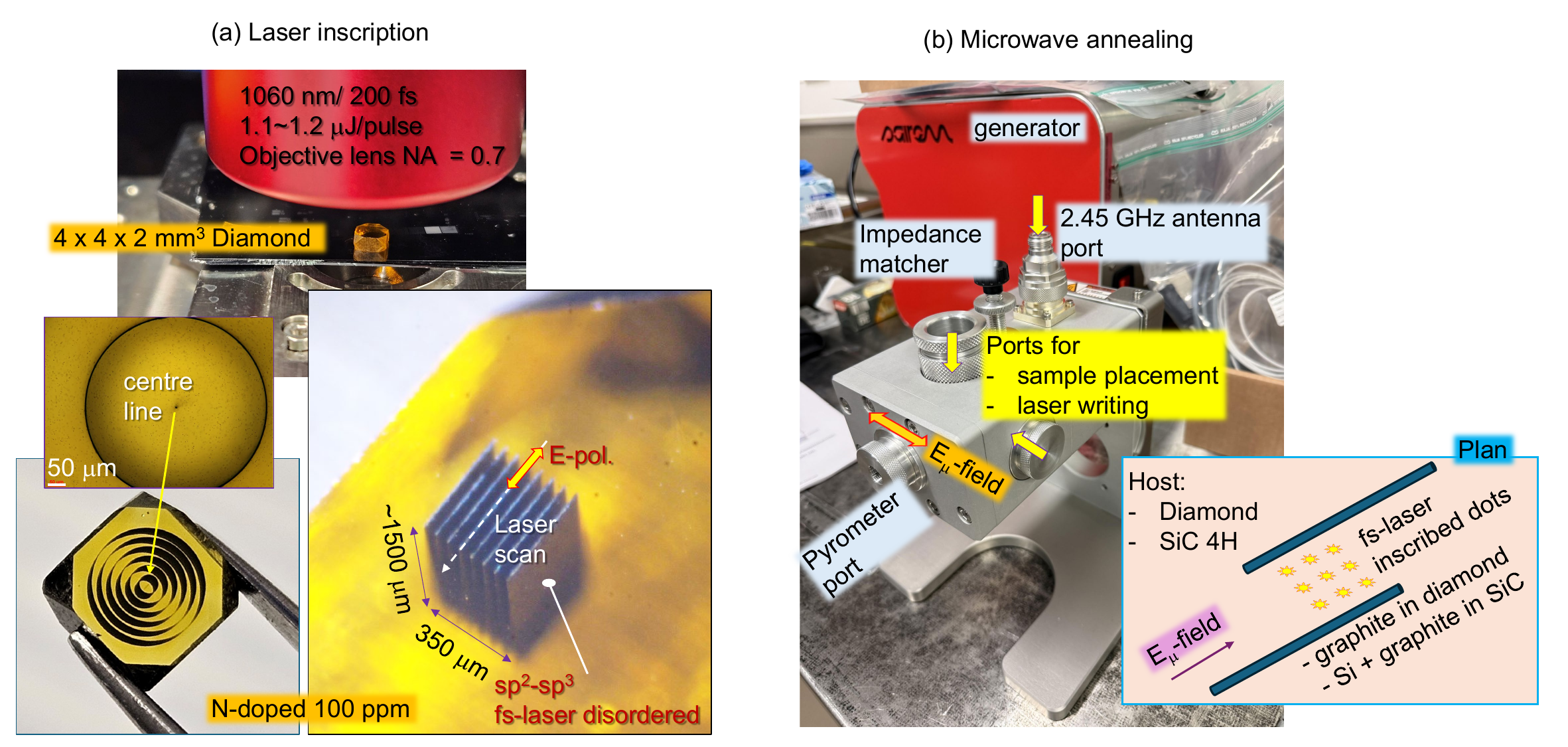}
\caption{Concept of fs-laser inscription of microwave absorbing antennas and nano-micro-volumes in the close proximity, which can be annealed in a microwave cavity. (a) Examples of THz Bessel beam generator (concentric grating) and polariser grating inscribed in a Nitrogen-doped (100 ppm) diamond. The width of the graphitised plane is $\sim 10~\mu$m. (b) Sairem microwave cavity setup (2.45~GHz, 200~W) for sample annealing. The inset shows the concept of graphitised antenna heaters with inscribed single-pulse structured nano-volume modifications for controlled annealing. Laser writing inside a microwave-heated sample is also feasible.
}\label{f-dream}
\end{figure*}
The localization physics above translates into engineering (Fig.~\ref{f-wafe}) whose common denominator is that direct write removes constraints that parallel lithography cannot.

\textbf{Stitching.} A central obstacle in large-area lithography is the stitching error between adjacent write fields, compounded by the cost of electron beam lithography (EBL)-made masks. Direct laser writing dissolves the stitching problem in two distinct ways~\cite{Juodkazis2026stitch}. (i) In a moving-beam moving-stage architecture the pattern is written continuously, avoiding field boundaries; combined with real-time surface tracking it scales to $\sim\!\SI{1}{m^2}$ masks. (ii) In
femtosecond direct write the written voxel is not a binary exposure but a multi-property material state (refractive index, birefringence, stress, crystallinity, conductivity); stitch-free fabrication is therefore a state-continuity problem, addressed by multi-dimensional holography and
polarisation-resolved (four-polarisation) diagnostics that close the fabrication
loop~\cite{Juodkazis2026stitch,Ryu2022,Honda2019}. Parallel holographic writing with spatial light modulators or metalens focal-spot arrays eliminates stitching by writing the whole pattern at once~\cite{Saha2019,Gu2025}.

\textbf{Solar cells and large-area masks.} Bessel-beam direct write of $\sim\!\SI{30}{}$--$\SI{40}{nm}$ alumina etch masks enables photonic-crystal light-trapping textures on Si, targeting above-Lambertian absorption~\cite{Katkus2024,Maksimovic2022,Hsieh2020} and world-record-class cell efficiencies~\cite{Haase2018}, with intentional non-ablated support pillars allowing undercut-tolerant masks over panel-scale ($\sim\!\SI{1}{m^2}$) areas written by polygon scanners at surface speeds approaching km/s~\cite{Henriques2024,Moskal2023}.

\textbf{Functional 3D devices.} The same toolbox writes domain structures in LiNbO$_3$ for nonlinear optics~\cite{Juodkazis2006LN}, conductive graphitic tracks and diffractive optics in diamond and wide-bandgap
semiconductors~\cite{Wu2023,Castelletto2021}, hyper-doped Si regions ($\sim\!10^{20}\,\mathrm{cm^{-3}}$) for room-temperature single-electron transistors~\cite{Tabe2010}, and high-pressure phases via confined
microexplosion~\cite{Juodkazis2006,Rapp2015,Vailionis2011}. Polarisation control
of ablation via the normal $E$-field component~\cite{Iwase2009} and in-situ THz
emission diagnostics of the focal currents~\cite{Huang2024} make these processes
both manufacturable and observable (in situ monitored and controlled).

\section{Outlook hypothesis: laser-seeded microwave volumetric deposition} \label{sec:paradigm}
\blue{This closing section is deliberately forward-looking: it is the hypothesis-generating element expected of an 
review--perspective, it builds directly on the absorber-defined localisation that is the review's central theme (the laser defines where while a second, longer-wavelength source supplies energy there). We subject it to a quantitative feasibility analysis below, which identifies heat conduction as a severe constraint rather than claiming a worked scheme.} We now develop the central hypothesis of this perspective. The threshold effect and the dipole/ENZ mechanism localise where energy is deposited to a tens-of-nm seed. Independently, microwave heating is the canonical volumetric and selective energy source in materials processing: just as a kitchen microwave oven deposits energy by shaking water molecules throughout the entire volume of the food rather than at its surface, a cavity field dissipates power only in lossy regions while leaving low-loss hosts cold~\cite{Bykov2001,Rybakov2013,Ravi2026}. The deposition is therefore intrinsically volumetric with no skin-depth or surface limitation at these scales, which is exactly what lets it fill a buried nanoseed uniformly.

We propose to combine the two: a femtosecond pulse, focused to a wavelength-scale spot ($d\sim\lambda\sim\SI{1}{\micro\meter}$)
inside a transparent sample, ionises a nanoscale seed at the tip of the Gaussian intensity profile by the threshold effect (Fig.~\ref{f-dream}). The sample sits inside a microwave cavity whose field is non-resonant and hence transparent to the pristine host, but couples strongly to the laser-defined lossy seed (or inscribed pattern). The microwave then supplies volumetric energy preferentially to the laser-defined structure and modified nano-volume (Fig.~\ref{f-wafe}, \ref{f-dream}). Whether this can sustain a localised hotspot against heat conduction is the central question, examined quantitatively and found restrictive as discussed next.

\subsection{Microwave coupling to the laser-defined nanovolume}\label{sec:mw}

Once the seed exists, it carries free or defect-band carriers and hence a finite conductivity $\sigma=n_e e^2/(m^\ast\nu)$, equivalently a loss $\varepsilon_{\rm in}''=\sigma/(\varepsilon_0\omega_{\rm mw})$ at the microwave frequency $\omega_{\rm mw}=2\pi f$, while the host has a negligible loss tangent. Three length-scale separations make the coupling clean:
\begin{enumerate}
\item \emph{Quasi-static.} $\lambda_{\rm mw}\sim\SI{1}{}$--$\SI{10}{cm}\gg a$, so
the cavity field $\bm E_{\rm mw}$ is uniform across the seed.
\item \emph{Field penetration.} The skin depth
$\delta=\sqrt{2/(\mu_0\sigma\omega_{\rm mw})}$ is macroscopic: for $\sigma=\SI{1e3}{S/m}$ and $f=\SI{2.45}{GHz}$, $\delta\approx\SI{320}{\micro\meter} \gg a$, so the field fully penetrates and heats the seed uniformly.
\item \emph{Per-volume selectivity.} The deposited power \emph{density} is larger in the seed than in the host by $S=\varepsilon_{\rm in}''/\varepsilon_h''\gtrsim 10^4$-$10^5$. This does \emph{not} mean the seed absorbs most of the \emph{total} power: integrated host and cavity-wall absorption scale with their (vastly larger) volumes, so selective \emph{total} heating of one seed requires many seeds, a
microwave near-field concentrator of sub-$\mu$m mode volume, or strong suppression of host/cavity loss.
\end{enumerate}
\noindent The internal field of the quasi-static sphere is $\bm E_{\rm int}=[3\varepsilon_h/(\varepsilon_{\rm in}+2\varepsilon_h)]\bm E_{\rm mw}$, and the time-averaged Joule (volumetric) power density is
\begin{equation}
\langle p_v\rangle=\tfrac12\,\sigma\,|\bm E_{\rm int}|^2
=\tfrac12\,\omega_{\rm mw}\varepsilon_0\varepsilon_{\rm in}''\,|\bm E_{\rm int}|^2.
\label{eq:joule}
\end{equation}
For a moderately conducting (impedance-matched) seed, $\varepsilon_{\rm in}'\approx\varepsilon_h$ and
$\bm E_{\rm int}\approx\bm E_{\rm mw}$, giving the working expression
\begin{equation}
\langle p_v\rangle\simeq\tfrac12\,\sigma\,E_{\rm mw}^2
\label{eq:pv}
\end{equation}
deposited uniformly throughout the seed volume $V=\tfrac43\pi a^3$. The localisation is set by $a$ (tens-of-nm) whereas the microwave wavelength is $\sim\!\SI{12}{cm}$: the energy is deposited into a region $\lambda_{\rm mw}/a\sim 5\times10^{6}$ times smaller than the wavelength sub-wavelength deposition by more than six orders of magnitude.

\subsection{Thermal runaway: sharpening toward single nanometers} \label{sec:runaway}

The conductivity of the seed rises with temperature in semiconductor-like fashion, $\sigma(T)=\sigma_0\exp(-E_a/2k_BT)$, with $\sigma_0$ a prefactor, $E_a$ the activation energy, $k_B$ Boltzmann's constant and $T$ the temperature. Substituting into Eq.~(\ref{eq:pv}) gives $\partial\langle p_v\rangle/\partial T>0$: hotter sub-regions absorb more, a positive feedback that is the well-known thermal-runaway/selective-absorption instability of microwave processing~\cite{Bykov2001,Rybakov2013}. Microwave high-temperature annealing was demonstrated in formation of nanoscale gradient P-doping of Si beyond the thermodynamical solubility~\cite{aplmic}.

Within the dipole dose profile, Eq.~(\ref{eq:r6}), the central maximum runs away first, contracting the actively heated zone, in principle toward smaller scales; but lattice heat diffusion (length $\sqrt{D\tau}$, thermal diffusivity $D$) imposes a stringent floor that\blue{, as Sec.~\ref{sec:numbers} shows,}
dominates at these parameters. Whether higher frequency helps is condition-dependent: at fixed conductivity $\sigma$, $\langle p_v\rangle=\tfrac12\sigma E^2$ is frequency-independent; at fixed $\varepsilon''$, $\sigma=\omega_{\rm mw}\varepsilon_0\varepsilon''$ rises with
$\omega_{\rm mw}$, but defect-relaxation and hopping losses are dispersive and roll
off at high frequency. A \SIrange{24}{30}{GHz} gyrotron source~\cite{BykovGyrotron2018}
thus helps only under a fixed-$\varepsilon''$, fixed-field assumption that must be
checked against the loss model. Deliberately driven, this runaway is also an opportunity rather than a nuisance: a self-localising,
microwave-sustained hotspot seeded by the laser could assist the dicing and
wafering of hard dielectrics and crystals: sapphire, SiC, diamond and optical
glasses where purely optical cutting is slow and the kerf hard to control, with
the microwave supplying the bulk of the separation energy volumetrically along the
laser-defined plane (Fig.~\ref{f-wafe}).

\subsection{Seed loss from point defects}\label{sec:defectloss}

How large is the seed loss in practice, and what sets the conductivity used below? The microwave conductivity is $\sigma=\omega_{\rm mw}\varepsilon_0\varepsilon''_{\rm in}$, so the range $\sigma=1$--$\SI{100}{S/m}$ corresponds to $\varepsilon''_{\rm in}\approx7$-$700$
at \SI{2.45}{GHz} (where $\omega_{\rm mw}\varepsilon_0=\SI{0.136}{S/m}$). Pristine
silica is essentially loss-free here ($\tan\delta\sim10^{-4}$, $\varepsilon''\sim4\times10^{-4}$, $\sigma\sim\SI{5e-5}{S/m}$), so the seed/host
contrast can exceed $10^4$. The laser-written core, by contrast, is saturated with
point defects: oxygen vacancies, $E'$ silicon dangling bonds, non-bridging-oxygen
and oxygen-interstitial (peroxy) centres~\cite{LoPiccolo2021} at densities
\blue{$n_d\sim10^{19}$-$10^{21}\,\mathrm{cm^{-3}}$. The upper decade corresponding to heavily exposed, near-breakdown volumes. The number densities of individual defect species in irradiated and as-drawn silica are catalogued in refs.~\onlinecite{Skuja1998,Pacchioni2000,LoPiccolo2021} and the local density in a laser-written core depends strongly on exposure dose and annealing history}. Two channels make these lossy at GHz. (i) Debye relaxation of defect dipoles contributes $\Delta\varepsilon'=n_d p^2/(3\varepsilon_0 k_BT)$; with a defect dipole
$p\sim e\times\SI{0.1}{nm}$ this gives $\Delta\varepsilon'\sim1$ at
$10^{21}\,\mathrm{cm^{-3}}$, peaking in $\varepsilon''$ near $\omega_{\rm mw}\tau\!=\!1$. (ii) \emph{Thermally activated hopping}/small-polaron conduction between defect sites is the mechanism shown to raise the GHz dielectric loss of oxygen-vacancy- and Ti$^{3+}$-rich titania-silica glass by orders of
magnitude~\cite{Zhang2025}, and to make oxygen-deficient oxides strong microwave
absorbers. Together these place $\varepsilon''_{\rm in}\sim1$--$10^2$ (hence
$\sigma\sim0.1$-$\SI{10}{S/m}$) within reach for a defect-saturated seed, rising
further if residual free carriers survive the microwave duration. 
A direct measurement of $\varepsilon''_{\rm in}(n_d)$ for laser-written volumes (Fig.~\ref{f-dream}) is the key input needed for precise estimates. It also provides a real-time feedback of material modifications during annealing.

\blue{\textbf{A practical measurement strategy for $\varepsilon''_{\rm in}(n_d)$.} The measurement is feasible with standard microwave metrology, provided the small modified-volume fraction is respected. A high-$Q$ cavity-perturbation or split-post dielectric-resonator measurement at 1--10\,GHz compares a reference wafer with an identical wafer carrying a laser-written grating of known volume fraction $\phi$ (e.g.\ $10^{3}$--$10^{5}$ written planes, $\phi\sim10^{-6}$--$10^{-4}$): the frequency shift and $Q$ degradation yield the composite $\Delta\varepsilon'$ and $\Delta\varepsilon''$, from which the intrinsic $\varepsilon''_{\rm in}$ follows by effective-medium inversion. Because $\Delta(1/Q)\approx\phi\,\varepsilon''_{\rm in}/\varepsilon_h$, resolving $\varepsilon''_{\rm in}\sim1$ at $\phi\sim10^{-5}$ requires $Q\gtrsim10^{6}$, demanding but routine in sapphire-loaded or superconducting cavities. The defect density $n_d$ in the same samples is quantified independently by electron spin resonance (EPR) and optical absorption of the known silica centres~\cite{Skuja1998,Pacchioni2000,Wang2016}. The written volumes can be re-measured after stepwise furnace or microwave annealing (Fig.~\ref{f-dream}), giving $\varepsilon''_{\rm in}$ versus $n_d$ on one sample set, and simultaneously testing the real-time-feedback idea above. The main experimental challenges are separating surface/adsorbate losses from bulk ones (addressed by measuring before/after writing on the same wafer) and the temperature dependence of the host loss during annealing runs. Feasibility of defect heating by microwaves is evaluated in Sec.~\ref{sec:numbers}.}

\section{Conclusions and outlook}\label{sec:outlook}
Three messages run through this review. First, the highest-resolution energy deposition is achieved not by shrinking the wavelength but by localising the absorber. The threshold effect [Fig.~\ref{f-Heis}(b)], the interface boundary conditions, and the dipole/ENZ near-field with its $I\propto r^{-6}$ tail (Figs.~\ref{fig:inscribe} and \ref{fig:gauss3d}) confine the deposited energy to a tens-of-nm seed whose size is set by the laser nonlinearity and the seed itself, not by $\lambda$.
\blue{Near-threshold exposure reaches the tens-of-nanometre scale and the $\sim\!\SI{10}{nm}$ modification scale when combined with the seed near-field mechanisms of Sec.~\ref{sec:dipole}.}
\blue{The final feature size further depends on the post-pulse material response: heat diffusion and quench rate, Eq.~(\ref{eq:wfinal}), which favours low-diffusivity hosts and confinement geometries.} This near-field route is intrinsically wavelength-insensitive, a conclusion that the latest surveys of ultrafast-laser nanofabrication~\cite{Lin2025,Stoian2023} and of resonant ENZ light-matter interaction~\cite{Xie2025} independently reinforce\blue{, and that holds whether the low-permittivity seed is created electronically (ENZ/Die-Met transit) or mechanically (cavitation/rarefaction, Sec.~\ref{sec:alt})}.

Second, polarisation controls the directionality and anisotropy of the deposition. The quasi-static near-field and the normal-$\bm D$/$\nabla\varepsilon$ boundary enhancement fix the writing direction, the finest lines are drawn with $\bm E$ perpendicular to the scan (Fig.~\ref{fig:polar}) and the orientation of the inscribed modification down to the single-defect scale (Fig.~\ref{fig:inscribe}). Together, the threshold effect and polarisation-steered near-field deposition constitute a maskless 2D and 3D direct-write nanolithography operating at the $\sim\!\SI{10}{nm}$ scale, already practised as the nano-pen on surfaces and the nano-knife in the bulk [Figs.~\ref{f-Heis}(c,d) and \ref{f-wafe}]. \blue{To keep claims and goals distinct: 10--20\,nm features, the polarisation-perpendicular writing rule, record-aspect-ratio dicing and polarisation-correlated single-emitter creation in h-BN are experimental results~\cite{Li2020,Li2024,Wang2024}.  Deterministic single-defect placement on demand, single-bond selectivity and layer-by-layer electrostatic surface removal remain research goals, and the mechanism attribution of the writing rule awaits dedicated polarisation-angle-resolved studies (Sec.~\ref{sec:polar}).}

These principles are not merely theoretical. A clean, benign read-out of the ENZ energy-deposition physics is the multiphoton-polymerisation voxel, whose abrupt
lateral-growth acceleration at the diffraction limit we analyse in Appendix~\ref{app:enz}: below the focal-spot size the Gaussian threshold sets the width, while above it the ionised \blue{core , plausibly, though not yet provably, an ENZ-class volumetric absorber} takes over and the two voxel boundaries widen independently.
The same maskless, 3D-capable physics carries an economic message that is increasingly strategic: as set out in Appendix~\ref{app:econ}, a $\sim\!\$1$M femtosecond writer can jump-start sub-10\,nm and 3D-photonic prototyping in any economy without access to the \$100M-class EUV steppers held by a handful of fabs, a route to
\emph{sovereign} nanofabrication capability that matters acutely amid today's fragile, interruption-prone supply chains.

Third, and finally, the laser-seeded microwave hypothesis extends the central logic to its conclusion. The microwave is a volumetric, molecular-scale energy-delivery source: it couples to the very dipoles, defects and free carriers that the laser creates, and so realises guided energy deposition at the nano-/molecular scale. The laser defines where (a tens-of-nm, ENZ-defined volume) and the cavity supplies energy preferentially there (per unit volume), at a scale more than six orders of magnitude below its own wavelength (Fig.~\ref{f-wafe}). We present this as a hypothesis, not a worked scheme: \blue{the feasibility analysis of Sec.~\ref{sec:numbers} shows that a single isolated seed cannot be held hot ($\Delta T\approx\SI{7}{mK}
$ at the assumed parameters, worsening as $a^{2}$),} so any realisation must rely on field concentration, dense seed ensembles, sub-nanosecond pulsing, or a threshold-bias role rather than primary continuous heating. Future work must quantify the seed conductivity and its lifetime, the cavity-sample coupling, and the balance of microwave power against conductive loss\blue{; a concrete measurement path for $\varepsilon''_{\rm in}(n_d)$ is given in Sec.~\ref{sec:defectloss}}. For fs-laser-inscribed macroscopic patterns of defects and seed-absorber antennas, the usual antenna arguments of coupling and microwave-field alignment would guide the localised heating inside the volume of the sample at the closest micro-nano proximity to the antenna. Such laser-inscribed microwave antennas would provide controlled thermal annealing of the neighbouring nano-/micro-scale structures and defects, from within the volume, beyond what currently available flash-annealing~\cite{flash} approaches deliver from the outside.

\small\begin{acknowledgments}
The authors acknowledge discussions with Eugene~G.~Gamaly, Vladimir~T.~Tikhonchuk, Hong-Bo Sun, Zhenze Lee, Lei Wang, Sajeev~John,
Leonid~V.~Zhigilei. 
Support from the ARC and the Vilnius University Laser Research Center is gratefully acknowledged. S.J. is grateful to Gintas \v{S}lekys and Workshop of Photonics Ltd., Lithuania for technology transfer project and first high-power femtosecond fab station at Swinburne Uni. This research was funded by ARC DP240103231 and Swinburne's Proof-of-Concept ``Silicon Carbide Single Photo Emitter Fabrication'' grants.
\end{acknowledgments}

\section*{Author Declarations}
The authors have no conflicts to disclose.

\section*{Data Availability}
Data sharing is not applicable to this review article as no new data were created
or analyzed.

\appendix
\setcounter{figure}{0}\setcounter{equation}{0}
\setcounter{section}{0}\setcounter{equation}{0}
\makeatletter
\renewcommand{\thefigure}{A\arabic{figure}}
\renewcommand{\theequation}{A\arabic{equation}}
\renewcommand{\thesection}{A\arabic{section}}

\section{ENZ onset at the focus during multiphoton polymerisation}\label{app:enz}

Multiphoton polymerisation (MPP) is a clean model system in which the transient ENZ/Die-Met onset can be read directly from the written voxel: a polymerisable resin is transparent and low-loss until the focal ionisation drives its
\emph{instantaneous} permittivity down. Using a broadly tunable source
(\SIrange{700}{1300}{nm}) with pulse-duration and dispersion control, 
the lateral and longitudinal voxel growth
and the dynamic fabrication window (DFW, the damage-to-polymerisation threshold
ratio) across the two- and three-photon bands was measured~\cite{Samsonas2023}. Two observations bear directly on the present picture: (i) the voxel growth shows an abrupt change in its power (fluence) dependence; and (ii) the DFW shrinks strongly toward longer wavelength,
from $\approx\!29$ at \SI{700}{nm} to $\approx\!1$ at \SI{1250}{nm}. Both are
interpreted as a consequence of a change in the instantaneous refractive index
(plasma/ENZ formation) that lowers the polymerisation, but not the
damage threshold~\cite{Samsonas2023}.

This can be sharpened with the dipole/ENZ mechanism of the main text. Below the
diffraction-limited focal-spot size, the modified lateral radius follows the
threshold-on-Gaussian law of Eq.~(\ref{eq:threshold}),
\blue{$a\simeq(w_0/\sqrt{2})\sqrt{\ln(E/E_{\rm th})}$}
: the voxel samples only the Gaussian
tail above threshold, so it widens logarithmically, i.e. slowly and traces the beam profile at the true nanoscale. Once $a$ approaches $w_0$, however, the ionised core fills the high-intensity focal volume and the Drude free carriers turn it into a volumetric absorber: the instantaneous permittivity falls toward ENZ, the internal field is enhanced [Eq.~(\ref{eq:internal3D})]. The absorbed power $\propto\varepsilon''|\bm E_{\rm int}|^2$ now grows with the carrier density the pulse itself creates. The lateral size therefore accelerates with pulse energy rather than saturating as a passive Gaussian-threshold profile would. The abrupt change in the power dependence is seen in MPP. In short, below $w_0$ the beam sets the size; above $w_0$ the absorber does.

\blue{\textbf{Status of this interpretation.} We emphasise, as a matter of scientific hygiene, that a change in the slope of feature size versus pulse energy does not by itself establish a transient ENZ state. It establishes a change of growth regime. The ENZ/volumetric-absorber reading is a possible explanation, favoured here because it follows from the same equations [Eqs.~(\ref{eq:threshold}), (\ref{eq:internal3D})] that account for the nano-pen and nano-knife, but at least three alternatives must be excluded by data: (i) radical/heat diffusion beyond the exposed volume (proximity effect), which also accelerates widening at high dose; (ii) cumulative thermal effects at high repetition rate; and (iii) focal aberrations that reshape the intensity distribution at high energy. 
}

The wavelength trend follows from the same physics. Free-carrier (Drude) absorption
scales as $\alpha_{\rm fc}\propto\lambda^2 n_e$, while the critical density at which
$\varepsilon'\!\to\!0$ scales as $n_c\propto\omega^2\propto\lambda^{-2}$; both make the ionised volume absorb more strongly, and reach ENZ at a lower carrier density, at longer wavelength. A focus at \SI{1100}{nm} therefore couples energy into its ionised core more efficiently than one at \SI{800}{nm} ($\alpha_{\rm fc}$ larger by $\sim\!(1100/800)^2\approx1.9$), giving a steeper
lateral growth and a correspondingly narrower fabrication window; 
the DFW collapse toward longer wavelength reported~\cite{Samsonas2023}. The
polymerisation voxel is thus a benign, direct read-out of the same ENZ
energy-deposition physics that, in harder materials, drives modifications discussed as the nano-pen (nanolithography)~\cite{Li2020} and
nano-knife (dicing and tomo-slicing or wafering)~\cite{Li2024,Li2026}.

\subsection*{Independent growth of the voxel boundaries at the diffraction limit}

A complementary and, we argue, likely contribution to the accelerated lateral
widening of Appendix~\ref{app:enz} is purely geometric. While the modified region is smaller than the focal spot, its two transverse sides lie within one near-field range of each other (the dipole field decays over $\sim\!a$), so they act as a single, coupled low-$\varepsilon$ core with one effective deposition centre. Once the voxel grows to the diffraction-limited focal-spot size, however, the two lateral boundaries become separated by more than that near-field length and decouple: each is now an independent permittivity step that carries its own normal-$\bm D$/$\nabla\varepsilon$ enhancement [Eqs.~(\ref{eq:Dnorm}) and
(\ref{eq:Esingular})], exactly as the two walls of a 2D groove do in the nano-pen.
The two fronts then deposit energy and advance independently of the voxel
centre, so the polymerised line widens from both edges at once. This doubling and
decoupling of the lateral growth channels sets in precisely at the focal-spot size
and provides a natural, geometric reason why the side widening accelerates
there rather than saturating, consistent with the abrupt change in the lateral
voxel dynamics
~\cite{Samsonas2023}. A falsifiable
corollary is that, deep in this regime, the line cross-section should evolve from
the single Gaussian-like core found below the diffraction limit toward a flat-bottomed or twin-lobed profile bounded by the two outward-moving edges.

\section{Feasibility estimates}\label{sec:numbers}

Table~\ref{tab:feasibility} collects representative power densities. Taking a cavity field $E_{\rm mw}=\SI{1e6}{V/m}$ ($\SI{10}{kV/cm}$, attainable from a modest $\sim\!\SI{100}{W}$ source in a high-$Q$ (cavity quality factor $Q$), field-concentrating (re-entrant) cavity), a seed of radius $a=\SI{25}{nm}$ ($V\approx\SI{6.5e-23}{m^3}$) and $\sigma=1$--$\SI{100}{S/m}$, Eq.~(\ref{eq:pv}) gives $\langle p_v\rangle=5\times10^{11}$--$5\times10^{13}\,\mathrm{W\,m^{-3}}$ and a deposited power per seed $P=\langle p_v\rangle V\sim10^{-11}$--$10^{-9}\,$W.

Heat transport is the binding constraint in its simplest form argues against single-seed heating. Indeed, a \SI{25}{nm} seed is far from adiabatic: with thermal diffusivity $D\sim\SI{1e-6}{m^2/s}$, the diffusion time is $\tau_{\rm diff}\sim a^2/D\sim\SI{0.6}{ns}$, so heat leaves the seed in nanoseconds, and over a \SIrange{0.2}{20}{ms} microwave dwell it spreads $\sqrt{D\tau}\sim\SI{14}{}$--$\SI{140}{\micro\meter}$, thousands of seed radii.
In steady state the temperature rise at the surface of a continuously heated sphere is $\Delta T=\langle p_v\rangle a^{2}/(3\kappa)$ (the centre is $1.5$ times hotter for equal conductivities):
for $\sigma=\SI{100}{S/m}$ and $a=\SI{25}{nm}$ this is only $\Delta T\approx\SI{7}{mK}
$ (silica thermal conductivity $\kappa\sim\SI{1.4}{W/m/K}$), and the size scaling is unfavourable. The absorbed power falls as $a^{3}$ while conductive loss falls only as $a$, so $\Delta T\propto a^{2}$ and smaller seeds heat less. Equivalently, the steady-state power needed to hold one seed at $\Delta T\sim\SI{1000}{K}$ is $P_{\rm cond}\sim4\pi\kappa a\,\Delta T\approx\SI{4e-4}{W}$,
five to seven orders of magnitude above the $3\times10^{-11}$--$3\times10^{-9}$\,W the cavity delivers per seed. Reaching $\SI{1000}{K}$ in a single \SI{25}{nm} seed with $\sigma=\SI{100}{S/m}$ would require $E_{\rm mw}\approx\SI{4e8}{V/m}$, a field enhancement of $\sim\!370$ over the assumed $\SI{1e6}{V/m}$. For a dense ensemble that fills a region of radius $R$ with volume fraction $f$, the same estimate $\Delta T=f\langle p_v\rangle R^{2}/(3\kappa)$ gives $R\approx\SI{9}{\micro\meter}$ for $f=1$ and $R\approx\SI{30}{\micro\meter}$ for $f=0.1$ at $\Delta T=\SI{1000}{K}$ (for $\sigma=\SI{100}{S/m}$), which is the length scale of the laser-inscribed planes and antennas of Figs.~\ref{f-wafe} and \ref{f-dream}.
A single isolated seed therefore cannot be held hot by the microwave at these parameters. The adiabatic times $\tau=U^\ast/\langle p_v\rangle$ in Table~\ref{tab:feasibility} (with $U^\ast\sim\rho c_p\Delta T\sim\SI{1e10}{J/m^3}$) are upper bounds that ignore conduction.

The proposal can close only if one or more of the following holds: (i) the microwave acts as a transient bias that lowers a threshold the laser already nearly meets, rather than as the primary heat source; (ii) field concentration raises $E_{\rm mw}$ (hence $P\propto E_{\rm mw}^2$) by several orders; (iii) a dense ensemble of seeds is heated collectively, so that the warm region has size $R\gg a$ and the per-seed conductive loss falls by $\sim\!a/R$. For the laser-inscribed planes of Fig.~\ref{f-wafe} with seed spacing $s\ll R$ this is the realistic operating mode, and macroscopic graphitised antennas (Fig.~\ref{f-dream}) are its deliberately engineered limit; or (iv) heating is pulsed on sub-$\tau_{\rm diff}$ timescales. We therefore present the microwave route as a hypothesis whose single-seed energetics are unfavourable, whose ensemble/antenna energetics are plausible, and whose quantitative inputs ($\varepsilon''_{\rm in}$, seed lifetime, cavity coupling) remain to be measured.
\begin{table}[t]
\caption{Representative power densities for laser-seeded microwave deposition. Seed radius $a=\SI{25}{nm}$, $E_{\rm mw}=\SI{1e6}{V/m}$, $U^\ast=\SI{1e10}{J/m^3}$. The times $\tau$ are the \emph{adiabatic} (no-conduction) upper bound; the conduction loss $P_{\rm cond}\sim\SI{4e-4}{W}$ at $\Delta T=\SI{1000}{K}$ greatly exceeds $P$ (see text).}
\label{tab:feasibility}
\begin{tblr}{colspec={lccc}, colsep=5pt, rowsep=1.2pt,
  hline{1,Z}={1pt,solid}, hline{2}={0.4pt,solid}}
$\sigma$ (S/m) & $\langle p_v\rangle$ (W\,m$^{-3}$) & $\tau=U^\ast/p_v$ & $P=p_vV$ (W)\\
1   & $5\times10^{11}$ & 20\,ms  & $3\times10^{-11}$\\
10  & $5\times10^{12}$ & 2\,ms   & $3\times10^{-10}$\\
100 & $5\times10^{13}$ & 0.2\,ms & $3\times10^{-9}$\\
\end{tblr}
\end{table}

\subsection{Experimental implementation and diagnostics}\label{sec:implementation}

The scheme requires (i) a high-NA femtosecond writer focusing inside a sample placed at an antinode of a microwave cavity (rectangular TE$_{10n}$ or cylindrical mode, $f=\SI{2.45}{GHz}$ magnetron or \SIrange{24}{30}{GHz} gyrotron), and (ii) synchronisation so that the cavity field is present during and after each pulse. Diagnostics already developed for USPL--matter interaction transfer directly: in-situ THz emission, which tracks currents and electron temperature in the focal volume~\cite{Huang2024}, four-polarisation imaging of near-field anisotropy below the diffraction limit~\cite{Ryu2022,Honda2019}, which directly visualises the polarisation-selected deposition of Sec.~\ref{sec:polar}, and plasma/optical breakdown spectroscopy. Polymerisable resins without photoinitiators provide a clean model system in which the ENZ onset is visible as a step in the polymerised cross-section~\cite{Samsonas2023}, allowing the laser-only and laser-plus-microwave regimes to be compared quantitatively. 

\section{Sovereign capability and the economics of maskless 3D direct write}\label{app:econ}

The scientific case above has an economic counterpart that may be decisive for
countries without access to extreme-ultraviolet (EUV) infrastructure. A High-NA EUV stepper now costs $\sim\!\$380$~million per tool (low-NA EUV
$\sim\!\$180$~million)~\cite{ASMLcost2024}, requires vacuum, reflective masks and a
$\sim\!10^3$-step planar flow, and is available to a mere handful of fabs. A
femtosecond direct-write station, by contrast, is a $\sim\!\$1$~million
instrument, two to three orders of magnitude cheaper and, being maskless and
3D-capable, can serve four distinct roles from one platform: (i) writing the
projection-lithography masks themselves, replacing slow electron-beam
mask-making (the nano-pen)~\cite{Li2020,Juodkazis2026stitch}; (ii) inscribing 3D photonic chips - buried waveguides and quantum emitters/colour centres in the bulk of glasses and crystals~\cite{Davis1996,Castelletto2018,Castelletto2021}; (iii) on-chip photonic wiring by two-photon
polymerisation~\cite{Kawata2001,Saha2019}; and (iv) sculpting 3D
micro-optical elements: lenses, Fresnel and diffractive optics, phase plates - out of glasses and crystals~\cite{Wu2023,Juodkazis2006LN}. This is the ``3D-printing-like'' democratisation that lets a laboratory in Australia, Lithuania or any non-EUV economy prototype sub-10\,nm and 3D devices at risk capital, in line
with sovereign-capability and supply-chain-resilience goals~\cite{Bhattacharya2020}.

A scaling argument shows why the maskless route wins precisely where modern devices are heading: 3D and photonic. A 2D projection stepper images one
layer per exposure, so an intrinsically three-dimensional structure of $L$
functional layers must be emulated by $L$
mask$\rightarrow$expose$\rightarrow$etch$\rightarrow$deposit cycles: cost and time
grow linearly in $L$ on top of a per-mask charge of order $\$10^6$. Direct
write instead lays the structure down volumetrically in a single pass, its
throughput set not by layer count but by the optical Moore's law: pulse repetition
rate times the number of parallel beams. At \SIrange{1}{1000}{MHz} rates with
$\sim\!10^2$ beams and polygon scanning at surface speeds approaching
km/s~\cite{Henriques2024,Moskal2023}, $10^{8}$-$10^{10}$ voxels per second are
written, so a $10^{10}$-voxel 3D photonic chip is completed in seconds to minutes
with no masks at all. \blue{Three caveats keep this comparison honest. First, the quantities compared are of different kinds: tool prices are vendor specifications, the km/s scan speeds and multi-beam counts are demonstrated engineering performance on specific tasks (large-area ablation of thin etch masks~\cite{Katkus2024,Henriques2024,Moskal2023}, parallel polymerisation~\cite{Saha2019,Gu2025}), while the voxel-per-second figures are idealised projections that assume every pulse writes a useful voxel. Second, a voxel rate is an upper bound on exposure throughput, not an estimate of device completion time: stage overheads, trajectory planning, in-situ metrology, development/etch steps where applicable, and, above all, yield qualification are excluded, and these dominate real fabrication schedules exactly as mask qualification dominates EUV insertion. Third, the comparison is task-specific: the materials and tasks underlying the direct-write figures are glasses, polymers and dielectric films patterned as photonic/mask structures, not transistor-grade silicon. 
} There is therefore a complexity cross-over. Below a few binary 2D layers, projection lithography is unbeatable for volume CMOS. Once $L$ becomes large and the geometry is genuinely three-dimensional, however, the sequential 3D writer can become faster and substantially cheaper. This regime includes photonic integrated circuits, whose market is projected to grow from approximately $\sim\!\$14$ billion in 2024 to $\sim\!\$26$ billion by 2030~\cite{GVRpic2024}, and which industry roadmaps increasingly pursue through 3D heterogeneous integration and co-packaged optics~\cite{IRDS2024}.

\blue{The division of labour between the two technologies is then clear and should be stated plainly. High-volume planar logic and memory remain the domain of EUV projection lithography: at $>10^{2}$ wafers per hour and $\sim\!10^{11}$ transistors per die, no sequential tool competes on throughput, and the mature yield-learning ecosystem is as decisive as the optics. Direct write offers the greatest value where EUV's strengths are irrelevant or its economics prohibitive: intrinsically 3D photonic devices (waveguide circuits, 3D photonic crystals), quantum-device prototyping (colour centres, single-emitter arrays, one-off chip variants where a mask set is unjustifiable), photomask writing and repair itself, low-volume/high-mix component manufacturing (micro-optics, diffractive elements, dicing/wafering of hard crystals), and large-area unique patterns (solar-cell light-trapping masks) where stitching, not pitch, is the binding constraint. Throughput, yield and device geometry decide the split: parallel projection wins when millions of identical planar layers amortise mask and qualification costs. Sequential direct write wins when the geometry is 3D, the volume is low, or the pattern is unique.}

Why, finally, do the two Moore's laws hold at all? Neither is a law of physics; both
are \emph{learning curves}. Empirically, cost falls as a power of cumulative
production (Wright's law) while production rises exponentially, and the two combine
into the exponential-in-time improvement we call Moore's law, a regularity
demonstrated across dozens of technologies and shown to be genuinely
forecastable~\cite{Nagy2013}. The e-Moore engine has ridden this self-reinforcing
investment-capability loop through parallel 2D scaling; the o-Moore engine rides the same loop in ultra-short-pulse laser power~\cite{Han2021}. The thesis of this review is that, as value migrates from planar transistor count to 3D photonic and quantum function, the o-Moore, sequential direct-write engine is poised to convert its exponential power growth into exponential fabrication throughput at a fraction of the capital cost.

\bibliographystyle{aipnum4-2}
\bibliography{main_R1}
\end{document}